\documentclass[12pt]{article}
\usepackage{cite}
\usepackage{showlabels}
\usepackage{enumerate}
\usepackage{ifpdf}
\usepackage{ae}
\usepackage[T1]{fontenc}
\usepackage[ansinew]{inputenc}
\usepackage{amsmath}
\usepackage{relsize}
\usepackage{amssymb}
\usepackage{comment}
\usepackage{graphicx}
\usepackage{color}
\definecolor{darkblue}{cmyk}{0.9,0.9,0,0}
\definecolor{darkgreen}{rgb}{0,0.55,0}
\usepackage[colorlinks=true,linkcolor=darkblue,citecolor=darkblue,urlcolor=darkblue]{hyperref}
\usepackage{epsfig}
\usepackage{epstopdf}
\usepackage{mathtools}
\usepackage{graphicx}

\usepackage{titlesec}
\usepackage{setspace}
\usepackage{tocloft}

\newenvironment{boenumerate}
  {\begin{enumerate}}
  {\end{enumerate}}

  {\begin{list}{}%
          {\setlength{\leftmargin}{#1} \setlength{\rightmargin}{#2}}%
          \item[]%
  }
  {\end{list}}
  
\makeatletter
\long\def\@makecaption#1#2{
  \vskip\abovecaptionskip
  \sbox\@tempboxa{{\captionfonts #1: #2}}
  \ifdim \wd\@tempboxa >\hsize
    {\captionfonts #1: #2\par}
  \else
    \hbox to\hsize{\hfil\box\@tempboxa\hfil}
  \fi
  \vskip\belowcaptionskip}
\makeatother

\def\it{\slshape}

\def\Ec{\mathcal{E}}
\def\mf{\mathfrak}
\def\n{\mf{n}}
\def\r{\mathbf{r}}
\def\R{\mathcal{R}}
\def\pst{\rho_{\,\rm ST}}
\def\A{\mathcal{A}}

\def\tree{{\rm tree}}
\def\cB{\mathcal{B}}
\def\gap{{\rm gap}}

\def\eps{\epsilon}

\def\bul{$\bullet$~~}

\def\Z{\mathbb{Z}}
\def\M{\mathcal{M}}

\def\1{{\rm 1-loop}}
\def\dd{\text{dDisc}}

\def\c{\cite}

\def\cA{\mathcal{A}}

\def\zb{\bar z}

\def\cN{\mathcal{N}}
\def\cA{\mathcal{A}}

\def\c{\cite}

\def\vs{\vskip .1 in}

\def\p{\partial}
\def\o{\over}
\def\g{\gamma}
\def\D{\Delta}
\def\rar{\rightarrow}

\def\eqr{\eqref}

\def\O{{\cal O}}
\def\ra{\rangle}
\def\la{\langle}
\def\ssec{\subsection}
\def\sssec{\subsubsection}
\def\sec{\section}
\def\i{\infty}
\def\foot{\footnote}
\newcommand{\es}[2] {\begin{equation} \label{#1} \begin{split} #2 \end{split} \end{equation}}
\newcommand{\e}[2] {\begin{equation} \label{#1} #2 \end{equation}}

\newcommand{\beq}{\begin{equation}}
\newcommand{\eeq}{\end{equation}}
\newcommand{\beqy} {\begin{eqnarray}}
\newcommand{\eeqy} {\end{eqnarray}}
\newcommand{\bsmat}{\begin{smallmatrix}}
\newcommand{\esmat}{\end{smallmatrix}}
\newcommand{\bmat}{\begin{matrix}}
\newcommand{\emat}{\end{matrix}}

\def\({\left(}
\def\){\right)}
\def\[{\left[}
\def\]{\right]}

\def\<{\langle}
\def\>{\rangle}

\def\a{\alpha}
\def\b{\beta}
\def\g{\gamma}

\def\D{\Delta}

\def\l{\lambda}

\def\t{\tau}

\def\vs{\vskip .1 in}

\usepackage{varioref}
\usepackage{makeidx}
\makeindex

\usepackage[english]{babel}
\usepackage{caption}
\usepackage{subfig}

\usepackage{tabularx}
 
\begin{document}

\thispagestyle{empty}

\renewcommand{\thefootnote}{\fnsymbol{footnote}}
\setcounter{page}{1}
\setcounter{footnote}{0}
\setcounter{figure}{0}
\begin{titlepage}
\null\hfill\begin{tabular}[t]{l@{}}
  CALT-TH 2019-018
\end{tabular}

\begin{center}


\vskip 5mm

{\LARGE \bf Growing Extra Dimensions in AdS/CFT
}
\vskip 0.5cm

\vskip 15mm
\centerline{ Luis F. Alday$^{D}$ and Eric Perlmutter$^{\M}$}
\bigskip
\centerline{\it $^{D}$ Mathematical Institute, University of Oxford,} 
\centerline{\it Woodstock Road, Oxford, OX2 6GG, UK}
\vs
\centerline{\it $^{\M}$ Walter Burke Institute for Theoretical Physics,}
\centerline{\it  Caltech, Pasadena, CA 91125, USA}

\end{center}

\vskip 2 cm

\begin{abstract}
\noindent What is the dimension of spacetime? We address this question in the context of the AdS/CFT Correspondence. We give a prescription for computing the number of large bulk dimensions, $D$, from strongly-coupled CFT$_d$ data, where ``large'' means parametrically of order the AdS scale. 
The idea is that unitarity of 1-loop AdS amplitudes, dual to non-planar CFT correlators, fixes $D$ in terms of tree-level data. We make this observation rigorous by deriving a positive-definite sum rule for the 1-loop double-discontinuity in the flat space/bulk-point limit. This enables us to prove an array of AdS/CFT folklore, and to infer new properties of large $N$ CFTs at strong coupling that ensure consistency of emergent large extra dimensions with string/M-theory. We discover an OPE universality at the string scale: to leading order in large $N$, heavy-heavy-light three-point functions, with heavy operators that are parametrically lighter than a power of $N$, are linear in the heavy conformal dimension. We explore its consequences for supersymmetric CFTs and explain how emergent large extra dimensions relate to a Sublattice Weak Gravity Conjecture for CFTs. Lastly, we conjecture, building on a claim of \c{ps}, that any CFT with large higher-spin gap and no global symmetries has a holographic hierarchy: $D=d+1$.

\end{abstract}

\end{titlepage}

\setcounter{tocdepth}{2}

\tableofcontents

\numberwithin{equation}{section}

\setcounter{page}{1}
\renewcommand{\thefootnote}{\arabic{footnote}}
\setcounter{footnote}{0}

 \def\nref#1{{(\ref{#1})}}

\sec{Introduction} 

This work aims to prove some fundamental aspects of the AdS/CFT Correspondence using bootstrap-inspired techniques at large $N$, and to extract new properties of strongly coupled large $N$ CFTs in the process. Our motivation comes from several directions:

\begin{boenumerate}
\item {\it Locality and sparseness in AdS/CFT}: Much has been written about the interplay between strongly coupled CFT dynamics and the hallmarks of gravitational effective field theory (e.g. \c{hpps, ps, showkimas, banks, fkap}). Famously, sub-AdS locality in an Einstein gravity dual requires a large gap to single-trace higher-spin operators in the CFT \c{hpps, cemz}: $\D_\gap\gg1$. But in {\it how many dimensions} is the bulk theory local? Large $\D_\gap$ is, despite occasional claims to the contrary, not the same thing as {\it sparseness} of the low-spin spectrum, another oft-invoked avatar of strong coupling. As we will explain, the answer to the previous question is directly correlated with the degree of sparseness. 

\item {\it String/M-theory landscape beyond supergravity:} What is the landscape of consistent AdS vacua? This question is old because it is challenging. The scales in the problem are the AdS scale $L$, the KK scale $L_\M$ (where $\M$ is some internal manifold), the Planck scale $\ell_p$, and (in string theory) the string scale $\ell_s$. Reliable bulk construction of scale-separated AdS vacua ($L_\M \ll L$) in string theory requires control at finite $\a'$ and, perhaps, at finite $g_s$. This is not currently possible without resorting to parametric effective field theory arguments, and/or assumptions about the structure of $\a'$ perturbation theory and/or backreaction of sources, which existing works all employ in some way.\foot{Some older works in the AdS/CFT context include \c{Polyakov:2004br, Klebanov:2004ya, Kuperstein:2004yk, Gadde:2009dj, Douglas:2010ic}. Some works from the swampland or string cosmological perspective include \c{kklt, Balasubramanian:2005zx,DeWolfe:2005uu,Acharya:2006zw,  McOrist:2012yc, Petrini:2013ika, Gautason:2015tig,Sethi:2017phn}.} 

\item{\it Bulk reconstruction from large $N$ bootstrap:} There has been recent progress in building up AdS amplitudes from large $N$ bootstrap, or bootstrap-inspired, methods. This is true both for ``bottom-up'' ingredients such as Witten diagrams \c{hpps, Alday:2014tsa, Alday:2016htq, Aharony:2016dwx, Li:2017lmh, Alday:2017gde, Giombi:2018vtc}, and top-down, complete amplitudes in string/M-theory at both genus zero \c{Rastelli:2016nze,Rastelli:2017udc,Rastelli:2017ymc,Chester:2018aca, Chester:2018dga,Binder:2018yvd,Caron-Huot:2018kta, Binder:2019jwn, Rastelli:2019gtj, Giusto} and genus one \c{Alday:2017vkk, Aprile:2017bgs, Alday:2017xua, Aprile:2017xsp, Aprile:2017qoy, Aprile:2018efk, Alday:2018pdi, Alday:2018kkw, Ponomarev:2019ltz}. It is natural to apply these insights to more abstract investigations of the AdS landscape. 

\end{boenumerate}
These topics invite many questions. We will answer the following one:
\begin{quotation} \noindent {\it Define $D$ as the number of ``large'' bulk dimensions, of order the AdS scale. Given the local operator data of a large $\D_\gap$ CFT to leading order in $1/c$, what is $D$?}
\end{quotation}

\noindent Unlike other questions in the realm of holographic spacetime, this is {\it not} possible to answer classically using a finite number of fields: consistent truncations exist. 

On the other hand, {\it quantum effects} in AdS {\it can} tell the difference between $D$ dimensions and $d+1$ dimensions. Our key idea is that AdS loop amplitudes are sensitive to $D$ because all $(d+1$)-dimensional fields generically run in loops. This philosophy, together with recent understanding of the structure of AdS loops imposed by unitarity \c{Aharony:2016dwx} and of the constructibility of CFT correlators from their double-discontinuity \c{LSPT,simon}, allows us to derive a positive-definite sum rule for $D$ in terms of tree-level data. Central to our approach is the flat space/bulk-point limit \c{Gary:2009ae, hpps, pen, Maldacena:2015iua, Hijano:2019qmi}, in which $D$ dimensions decompactify and the correlator becomes an S-matrix.

The sum rule is given in \eqr{main}, where $\b_{n,\ell}^\1$ is defined in \eqr{b1}. The formula \eqr{b1} holds for arbitrary $\D_\gap$, but the sum rule \eqr{main} holds at large $\D_\gap$. Its application efficiently proves statements about bulk emergence from large $N$, large $\D_\gap$ CFT that are widely believed to be true based on inference from the bulk side of tractable AdS/CFT examples. (See Sections \ref{sec2} and \ref{sec3} and Appendix \ref{appa}.) They include the following:
\begin{itemize}
\item  Degree-${\tt x}$ polynomial growth of single-trace degeneracies grows ${\tt x}$ large extra dimensions: $D=d+1+{\tt x}$. This is the converse of the inference of such growth from Weyl's law scaling of eigenvalue degeneracies of a transverse manifold $\M$ in an AdS$\times \M$ solution. 

\item For superconformal CFTs (SCFTs), we derive a ``characteristic dimension'' due to the existence of towers of Higgs or Coulomb branch operators. This places a lower bound on $D$ for SCFTs with these structures. The same is true for other non-R global symmetries, quantified in terms of the dimension of their irreps.

\item Requiring that AdS vacua are realized in critical or sub-critical string theory $(D\leq 10)$ or M-theory $(D\leq 11)$ bounds the asymptotic growth of planar CFT data at large quantum numbers. This implies quantitative bounds on which global symmetries with towers of charged operators may be geometrized at the AdS scale. 

\end{itemize}

Along the way, we make several remarks about emergent behavior at strong coupling. A particularly notable one is a new universal property of the OPE in planar CFTs, which we motivate and explore in Section \ref{sec4}: at leading order in $1/N$, the three-point coupling $C_{\phi pp}$ of two ``heavy'' operators $\O_p$, with $\D_p \propto p$ where $1\ll p \ll N^{\#>0}$, to a ``light'' operator $\phi$ with $\D_\phi \ll \D_p$ is {\it linear in $\D_p$}. This appears to hold generically. Applied to CFTs with string duals where $\O_p$ is taken to be a string state, this is a universality in the string scale OPE. We provide several arguments, and independent evidence from myriad CFT computations in the literature. While we arrived at this by way of arguments from 1-loop unitarity, this is a general property, independent of holography, that we expect to find wider applicability. We explain an implication for non-BPS spectra of SCFTs in Section \ref{nonbps}. In Appendix \ref{applinear} we sketch a worldsheet proof and discuss other consequences of the OPE linearity. This includes a connection between this OPE asymptotic, large extra dimensions, and a parametric form of the Sublattice Weak Gravity Conjecture for CFTs which we formulate in \eqr{caswgc}: the upshot is that not all symmetries in all large $c$, large $\D_\gap$ CFTs satisfy even a parametric form of the sLWGC, and whether this happens is closely related to large extra dimensions in the bulk. 

Finally, in Section \ref{sechhc} we apply lessons learned to formulate a ``Holographic Hierarchy Conjecture'' on the conditions required for prototypical bulk locality, $D=d+1$. This conjecture, like much of this paper, owes inspiration to the discussion of \c{ps}. It also makes contact with recent investigations into symmetry and UV consistency in quantum gravity \c{ov, fk, ho1, ho2}. We believe our intuition from the 1-loop sum rule to place the suggestions of \c{ps} on a firmer, and slightly modified, conceptual footing. 

\sec{Counting dimensions with 1-loop amplitudes}\label{sec2}
As stated in the introduction, we will present a formula from which one can derive $D$ from planar single-trace data in a dual CFT.

\ssec{1-loop sum rule}
We begin with some dimensional analysis. Consider a $D$-dimensional two-derivative theory of gravity coupled to low-spin matter ($\ell\leq 2$). Its scattering amplitudes admit a perturbative expansion in $G_N$. In this expansion, the connected four-point amplitudes of massless fields -- call them $A_D(s,t)$, where $s$ and $t$ are the two independent Mandelstam invariants -- take the form
\e{}{A_D(s,t) = G_N A_D^\tree(s,t) + G_N^2 A_D^\1(s,t) + \O(G_N^3)}
where $G_N$ is the (renormalized) $D$-dimensional Newton's constant. Suppose this theory admits a solution of the form AdS$_{d+1}\times \M$, where $\M$ is an internal manifold with dim$(\M)=D-d-1$. The solution need not be a direct product, but we will continue to use this simple notation. The amplitude $A_D(s,t)$ may be trivially expanded in powers of the AdS$_{d+1}$ Newton's constant by dividing by Vol($\M) \sim L_\M^{D-d-1}$. Call the resulting amplitude $A_{d+1}(s,t)$,
\e{}{A_{d+1}(s,t) \equiv{A_D(s,t)\o \text{Vol}(\M)}}
In order for this solution to be consistent with bulk effective field theory, $L_\M \sim L$, where $L$ is the AdS scale. Using the holographic dictionary, we can express $G_N$ in terms of $c$, the CFT central charge, to leading order, as
\e{}{G_N \sim {L^{D-2}\o c}}
Then expressing $A_{d+1}(s,t)$ in these variables, 
\e{Ad+1}{A_{d+1}(s,t) = {L^{d-1}\o c}A_{d+1}^\tree(s,t) + {L^{D+d-3}\o c^2}A_{d+1}^\1(s,t) + \O(c^{-3})}
By dimensional analysis, at high energy $s,t \gg1$,
\e{Alarges}{A_{d+1}(s\gg1,\theta) = {(L\sqrt{s})^{d-1}\o c}f_{d+1}^\tree(\theta) + {(L\sqrt{s})^{D+d-3}\o c^2}f_{d+1}^\1(\theta) + \O(c^{-3})}
where $\cos\theta = 1+{2t\o s}$ is kept fixed and $f_{d+1}^\tree(\theta)$ and $f_{d+1}^\1(\theta)$ are some angular functions. This expression must be reproduced by the flat space limit of an AdS$_{d+1}$ four-point amplitude, order-by-order in $1/c$. 

The AdS$_{d+1}$ amplitude is dual to a CFT$_d$ four-point function. For simplicity, we study a scalar four-point function of a single-trace primary operator $\phi$,
\e{}{\la \phi(0)\phi(z,\zb)\phi(1)\phi(\i)\ra =(z\zb)^{-\D_\phi}\A(z,\zb)}
We assume an orthonormal basis of single-trace operators on $\mathbb{R}^d$. To make contact with the dual AdS amplitude, let us write the $1/c$ expansion of the connected correlator as
\e{}{\A(z,\zb) = {1\o c}\A^{\tree}(z,\zb) + {1\o c^2}\A^\1(z,\zb) + \O(c^{-3})}
$\A(z,\zb)$ may be reconstructed from its double-discontinuity via the Lorentzian inversion formula \c{simon}, where we define the t-channel dDisc as the sum of the two discontinuities across the branch cut starting at $\zb=1$ (with $z$ fixed),
\e{}{\dd(\A(z,\zb)) \equiv {1\o 2}\text{Disc}_{\zb=1}^{\circlearrowleft}(\A(z,\zb)) + {1\o 2}\text{Disc}_{\zb=1}^{\circlearrowright}(\A(z,\zb))}
We would like to understand $\dd(\A^\1(z,\zb))$ and relate it to \eqr{Ad+1} in the flat space/bulk-point limit. 

In general, the t-channel exchange of a primary with twist $\tau\equiv \D-\ell$ contributes to $\dd(\A(z,\zb))$ as
\e{ddaction}{\dd(\A(z,\zb)) = 2C_{\phi\phi\tau}^2\sin^2\left(\pi\left({\tau-2\D_\phi \o 2}\right)\right)\left({z\zb\o (1-z)(1-\zb)}\right)^{\D_\phi}G_{\D,\ell}(1-z,1-\zb)}
where $G_{\D,\ell}(1-z,1-\zb)$ is the t-channel conformal block, normalized as 
\e{}{G_{\D,\ell}(1-z,1-\zb) \sim ((1-z)(1-\zb))^{\tau\o 2}}
in the t-channel limit $(z,\zb)\rar1$. What are the contributions to $\A^\1(z,\zb)$? There is renormalization of tree-level single-trace exchanges -- for example, (normalized) single-trace three-point coefficients $\la \phi\phi\O\ra$ can receive non-planar corrections -- but for the purposes of our discussion we can safely ignore these terms, as we will justify later. All other contributions are double-trace exchanges $[\phi\phi]_{n,\ell}$ and $[\O\O]_{n,\ell}$, where $\O$ is another single-trace operator; by large-$c$ counting, triple- and higher multi-traces don't appear in $\A(z,\zb)$ until $\O(c^{-3})$, i.e. 2-loop in AdS.\foot{$[\phi\O]_{n,\ell}$ and $[\O\O']_{n,\ell}$, with $\O\neq \O'$, may also appear in the 1-loop amplitude, but their three-point functions with $\phi\phi$ are allowed to be zero at $O(c^{-1})$ because there is no stress-tensor exchange in any channel of $\la \phi\phi\phi\O\ra$ or $\la \phi\phi\O\O'\ra$; this is as opposed to $\la \phi\phi[\O\O]_{n,\ell}\ra$, which must be nonzero at $\O(c^{-1})$ due to stress tensor exchange $\phi\phi \rar T \rar \O\O$. For our purposes, and given this lack of universality, we will only include $[\O\O]_{n,\ell}$ within this class, leaving the (straightforward) inclusion of the others implicit. Also, $\O$ can have spin, but because spin-dependence is not important for us, we are leaving implicit the different Lorentz representations that the double-traces ``$[\O\O]_{n,\ell}$'' can furnish.} The double-traces are spin-$\ell$ primaries defined schematically as
\e{}{[\phi\phi]_{n,\ell} \equiv \phi \square^n \p_{\mu_1}\ldots \p_{\mu_\ell}\phi - (\text{traces})}
where $n,\ell \in \Z_{\geq 0}$, and likewise for $[\O\O]_{n,\ell}$. In the $1/c$ expansion, its anomalous dimension $\g_{n,\ell}$ is
\e{}{\g_{n,\ell} = {1\o c}\g_{n,\ell}^{(1)} + {1\o c^2}\g_{n,\ell}^{(2)} + \O(c^{-3})}
and similarly for the squared OPE coefficients,
\e{}{a_{n,\ell} \equiv C_{\phi\phi[\phi\phi]_{n,\ell}}^2 = a_{n,\ell}^{(0)} + {1\o c}a_{n,\ell}^{(1)} + {1\o c^2}a_{n,\ell}^{(2)} + \O(c^{-3})}
$a_{n,\ell}^{(0)}$ are the squared OPE coefficients of Mean Field Theory (MFT). 

Let us first consider the contribution of $[\phi\phi]_{n,\ell}$ to $\dd(\A^\1(z,\zb))$. From \eqr{ddaction}, these are clearly proportional to squared anomalous dimensions of $[\phi\phi]_{n,\ell}$, where the overall coefficient is determined by
\e{}{\dd((1-\zb)^n \log^2 (1-\zb)) =4\pi^2 (1-\zb)^n}
The $[\O\O]_{n,\ell}$ terms are more interesting. Each $\O$ is a single-trace operator. There are two categories of $\O$. First, those with $\tau_{\O} \neq \D_\phi+\Z$ give rise to double-traces which do not mix with $[\phi\phi]_{n,\ell}$ and only appear for the first time in the $\phi\times\phi$ OPE at $\O(c^{-2})$, since $C_{\phi\phi [\O\O]_{n,\ell}} \sim c^{-1}$. (This scaling follows from the coincident limit of the connected four-point function $\la \phi\phi\O\O\ra$.) On the other hand, those with $\tau_{\O} = \D_\phi+\Z$ lead to double-trace mixing between $[\phi\phi]_{n,\ell}$ and the finite subset of $[\O\O]_{n',\ell}$ with $\D_\phi+n=\D_\O+n'$. These operators contribute in the same way as $[\phi\phi]_{n,\ell}$. The mixing leads us to define
\e{}{\la  (\g^{(1)}_{n,\ell})^2 \ra \equiv  \sum_\O \pst(\D_\O) \la\g^{(1)}_{n,\ell}(\O)\ra^2}
where $\pst(\D_{\O})$ is the single-trace density of states, and $\la\g^{(1)}_{n,\ell}(\O)\ra$ is the contribution to the anomalous dimensions due to mixing of $[\phi\phi]_{n,\ell}$ with $[\O\O]_{n',\ell}$.  Our notation is that $\la\g^{(1)}_{n,\ell}(\O)\ra=0$ when $\tau_\O \neq \D_\phi+\Z$. 

Therefore, modulo terms whose exclusion we explained earlier, the general 1-loop double-discontinuity $\dd(\A^\1)$ has the following conformal block decomposition:
\e{dd1}{\dd(\cA^{\1}(z,\zb)) = \left({z \zb\o (1-z)(1-\zb)}\right)^{\D_\phi}\sum_{n,\ell} \beta^\1_{n,\ell} a^{(0)}_{n,\ell}\,G_{n,\ell}(1-z,1-\zb)}
with coefficients
\e{b1}{\beta^\1_{n,\ell} \equiv 2\sum_\O \pst(\D_\O)\left({\pi^2\o 4} \la\g^{(1)}_{n,\ell}(\O)\ra^2 + \sin^2(\pi(\tau_{\O}-\D_\phi))\lVert C^2_{\phi\phi[\O\O]_{n,\ell}}\rVert\right)}
Our notation is that $a^{(0)}_{n,\ell}G_{n,\ell}(1-z,1-\zb)$ is the weighted conformal block for the exchange of a given double-trace operator $[\O\O]_{n,\ell}$ in the sum $\b^\1_{n,\ell}$ (i.e. that $\b^\1_{n,\ell}$ acts as a projector). For future convenience we have defined the MFT-normalized OPE data
\e{}{\lVert C^2_{\phi\phi[\O\O]_{n,\ell}}\rVert \equiv { C^2_{\phi\phi[\O\O]_{n,\ell}}\o a^{(0)}_{[\O\O]_{n,\ell}}}}
It is obvious that as $\O$ and $\phi$ become degenerate modulo integers, a contribution to the second term shifts to the first term. For conceptual reasons we have split off the two types of terms. In what follows, we sometimes refer to the first and second terms of \eqr{b1} as the ``integral'' and ``generic'' pieces, respectively.  

We now consider the flat space limit. The external local operators that define the correlator are located on the Lorentzian cylinder, boundary of AdS$_{d+1}$. By using suitable wavepackets we can focus onto a point in the bulk, effectively accessing (bulk) flat space physics, see  \c{Gary:2009ae, hpps, pen, Maldacena:2015iua}. At the level of the cross-ratios the relevant limit is $z-\bar z \to 0$, where the limit is taken after analytically continuing the Euclidean correlator to the Lorentzian regime of the scattering process. In this limit the correlator develops a bulk-point singularity. It can be seen that the leading singularity arises from intermediate operators with large scaling dimension, including double-traces with $n\gg1$. At large $n$, the leading order relation between the AdS radius $L$, the flat space center-of-mass energy $s$, and the double-trace twist $n$ is
\e{Lsn}{L \sqrt{s} \sim 2n}
On the other hand, the angular dependence of the flat space S-matrix is encoded in the spin dependence of the CFT data in the $n\gg1$ limit. For example, at tree-level, anomalous dimensions obey \c{Cornalba:2007zb}
\begin{equation}
\langle \gamma_{n \gg1,\ell}^{(1)}\rangle  \sim n^{d-1} f^{\rm tree}(\ell)
\end{equation}
where $f^{\rm tree}(\ell)$ is a function of spin which encodes the angular dependence of the tree-level S-matrix. At 1-loop, comparing the $\O(c^{-2})$ term of \eqr{Alarges} with \eqr{dd1}, and taking the flat-space limit, we read off the scaling of the 1-loop OPE data as $n\gg1$:
\e{main}{{\beta^\1_{n\gg1,\ell}}
\sim n^{D+d-3}f^\1(\ell)}
As before, $f^\1(\ell)$ encodes the angular dependence of the 1-loop S-matrix. 

Equation \eqr{main}, together with \eqr{b1}, is our central observation. The key feature is the $D$-dependence of the $n$-scaling. In addition, the coefficients $\b^\1_{n,\ell}$ are manifestly positive:
\e{}{\b^\1_{n,\ell} >0}
In fact, every term in the sum \eqr{b1} is positive. Therefore, knowledge of the tree-level single-trace OPE data in the $n\gg1$ regime may be used to read off $D$ using \eqr{main}. The scaling \eqr{main} followed from $\D_\gap \gg 1$, which allows us to probe only the $D$ large bulk dimensions. Positivity ensures the absence of cancellations. Note that taking the double-discontinuity was crucial: positivity is {\it not} an inherent property of the partial wave coefficients of $\A^\1$, but it {\it is} a property of $\dd(\A^\1)$.\foot{In comparing the scaling of dDisc$(\A^\1)$ to that of $A_{d+1}$, we are using the fact that $\A^\1$ can be constructed from dDisc$(\A^\1)$ \c{simon}; equivalently  \c{Alday:2017vkk}, that $A_{d+1}$ may be constructed via dispersion relations from its Disc($A_{d+1}$).}

We emphasize that we are using a {\it 1-loop} amplitude, of a single operator $\phi$, to derive the emergence of bulk spacetime from {\it tree-level} data. Note from \eqr{Alarges} that individual tree-level amplitudes are insufficient for this purpose: the behavior $\beta^\tree_{n\gg 1,\ell} \sim n^{d-1}$ is the well-known scaling of OPE data in a CFT$_d$ with $\D_\gap\gg1$, dual to Einstein gravity in AdS$_{d+1}$.\foot{This scaling is subleading to \eqr{main}, which justifies having dropped single-trace renormalization terms at 1-loop.} This reflects the existence of consistent truncations, valid only classically.

\ssec{Extra dimensions at large $\D_\gap$}
Equations \eqr{main} and \eqr{b1} lead to the following conclusions. Let's first focus on the generic piece of \eqr{b1}. In a typical chaotic CFT with an irrational spectrum of low-energy states, most contributions are of this type. To relate single-trace operator growth to $D$, we first note that at $\D_\gap\gg1$ for finite $\D_\O$, 
\e{c2gap}{\lVert C^2_{\phi\phi[\O\O]_{n,\ell}}\lVert\big|_{n\gg1} \sim n^{2(d-1)}f_\ell(\D_\O)}
for some nonzero $f_\ell(\D_\O)$.\foot{Henceforth we will not include the $\ell$-dependence in similar expressions, since we focus on the $n$-scaling. The symbol ``$\sim$'' captures the leading scaling in $n$, up to multiplicative factors which we suppress.} This just follows from graviton dominance at strong coupling, i.e. the large $\D_{\rm gap}$ condition. For a more detailed argument see Appendix \ref{appCa}. Now suppose that the single-trace degeneracy has asymptotic polynomial growth,
\e{pstx}{\pst(\D_\O\gg1) \sim (\D_\O)^{{\tt x}-1}}
Inserting this into \eqr{b1}, we want to extract the asymptotic growth at $n\gg1$. We must also account for the $\D_\O\gg1$ growth of the operator sum. Because the radial localization of AdS two-particle primary wavefunctions is determined by the total energy, dual to the conformal dimension $\D_{n,\ell} = 2\D_\O+2n+\ell+\g_{n,\ell}$, the parametric regime relevant for the computation is\foot{One can also examine explicit double-trace data as a function of $n$ and $\D_\O$, taking either one large with the other held fixed or taking a double-scaled limit of fixed $n/\D_\O$, whereupon one always finds the same scaling. A particularly relevant example is the universal contribution to anomalous dimensions due to stress-tensor exchange, $\g^{(1)}_{{n,\ell}}|_T$, which was derived explicitly in $d=4$ in \c{Alday:2017gde} and in general $d$ in \c{star1, star2}.} 
\e{regime}{1\ll n \sim \D_\O\ll \D_\gap}
Plugging into \eqr{b1} and approximating the sum as an integral up to $\D_\O\sim n$, we find the following $n$-scaling:
\es{}{\sum_\O \pst(\D_\O)  \sin^2(\pi(\tau_{\O}-\D_\phi))\lVert C^2_{\phi\phi[\O\O]_{n,\ell}}\lVert~ = {n^{2d+{\tt x}-1}\o 2(x+1)}+\O(n^{2d+{\tt x}-2})}
Therefore, we read off from \eqr{main} that
\e{d1x}{D=d+1+{\tt x}}
That is, polynomial growth $\int^{\D_*} d\D \,\pst(\D)\sim \D_*^{{\tt x}}$ implies the emergence of exactly ${\tt x}$ large bulk dimensions.

This is the converse of a basic fact about AdS$\times \M$ backgrounds in holography. Suppose $\M$ is a compact manifold with smooth boundary, of real dimension $\text{dim}(\M)$. Anticipating the holographic application, we parameterize the eigenvalues $\l$ of Laplace-Beltrami operators on $\M$ as $\l=(mL_\M)^2 \sim \D^2$ and the degeneracy as $\rho_\M(\D)$. Then Weyl's law states that 
\e{}{\int^{\D_*\gg1} d\D \,\rho_\M(\D) \sim {\text{vol}(\M)\o (4\pi)^{\text{dim}(\M)\o 2}\Gamma\left({{\text{dim}(\M)}\o 2}+1\right)}\,\D_*^{\text{dim}(\M)}}
Via $Z_{\rm AdS} = Z_{\rm CFT}$, one infers the polynomial growth of single-trace operators in the CFT. Our derivation from the CFT proves the converse: {\it every} large $N$, large $\D_{\rm gap}$ CFT with degree-${\tt x}$ polynomial growth of single-trace degeneracies grows ${\tt x}$ large bulk dimensions, i.e. the dimensions of $\M$. 

Now we turn to the integral piece of \eqr{b1}. The anomalous dimension receives a universal contribution from the presence of the stress tensor in the $\phi\times\phi$ OPE. Under a crossing transformation, t-channel stress tensor exchange maps to s-channel $[\phi\phi]_{n,\ell}$ exchange, with fixed anomalous dimension, $\g_{n,\ell}\big|_T$. This contribution scales at $n\gg1$ as
\e{gammaT}{\g^{(1)}_{n\gg1,\ell}\big|_T \sim n^{d-1}}
Now let us assume that there exists a single tower of scalar operators $\O_p$ with dimensions\foot{The generalization to spinning $\O_p$ just requires more technical aspects, e.g. multiple tensor structures \c{cppr}.}
\e{}{\D_p = \D_\phi+p-2~, ~\text{where}~ p=2,3,\ldots~.}
where $\O_2\equiv \phi$. We choose this convention to make contact with $\cN=4$ super-Yang-Mills (SYM), where $\O_p$ are superconformal primaries and $\O_2=\O_{\bf 20'}$. To compute \eqr{b1}, we need to resolve the double-trace operator mixing within the family
\e{}{[\phi\phi]_{n,\ell}~, ~ [\O_3\O_3]_{n-1,\ell}~,~\ldots~,~ [\O_{n+2}\O_{n+2}]_{0,\ell}~.}
This can be done by computing the family of {\it tree-level} correlators $\langle \phi\phi pp \rangle$. In particular, the t-channel exchange $\phi p \to p \to \phi p$ implies a contribution to tree-level anomalous dimensions $\la\gamma_{n,\ell}^{(1)}(p)\ra$. This exchange is manifestly proportional to the squared OPE coefficient $C^2_{\phi pp}$ at leading order in $1/c$, and therefore so is $\la\gamma_{n,\ell}^{(1)}(p)\ra$. In our flat space limit context, we want to evaluate this in the regime (cf. \eqr{regime})
\e{npregime}{1 \ll n\sim p \ll \D_\gap}
Based on computations in $d=2,4,6$ using explicit expressions for conformal blocks, we find
\begin{equation}\label{gammap}
\la\gamma_{n,\ell}^{(1)}(p)\ra\Big|_{1\ll n\sim p} \sim C^2_{\phi pp}\big|_{p\gg1}\times n^{d-3}
\end{equation}
The kinematic piece scales as $n^{d-3}$, but the overall scaling depends on the asymptotics of $C_{\phi pp}$. The derivation of \eqr{gammap} is given in Appendix \ref{appgamma}.

We can now put the pieces together. In the absence of the tower $\O_p$, the stress tensor contribution gives
\e{}{ \big(\g^{(1)}_{n\gg 1,\ell}\big|_T\big)^2 \sim n^{2(d-1)}~.}
By \eqr{main}, this implies $D=d+1$. This is the expected result: in a theory of gravity coupled to a finite number of fields, no large extra dimensions are required. Now we add the tower $\O_p$. Parameterizing
\e{cppp}{C_{\phi pp}\big|_{p\gg1} \sim {p^{1+{\a\o 4}}\o \sqrt{c}}~,}
where the $1/\sqrt{c}$ follows simply from large-$c$ factorization, we plug \eqr{cppp} and \eqr{gammap} into \eqr{b1}. Then \eqr{main} implies
\e{Dtowera}{D=d+2+\a}
It is interesting that this result depends on the asymptotics of $C_{\phi pp}$. If $C_{\phi pp}\big|_{p\gg1} \sim p$, then
\e{Dtower}{D=d+2}
A single extra dimension has grown. $C_{\phi pp}$ is, in fact, proportional to $p$ when $\phi$ is a scalar in the stress tensor multiplet of an SCFT: the action of supersymmetry relates $C_{\phi pp}$ to $C_{Tpp}\propto {\D_p/ \sqrt{c}}$, which is fixed by the conformal Ward identity. $C_{\phi pp}\propto p$ also holds when $\phi$ is the Lagrangian operator, by a conformal perturbation theory argument of \c{costa}. The above computation suggests that $\a=0$ is natural, in the sense that a single tower of $\O_p$ contains an $S^1$ worth of operators. In Section \ref{sec4}, we will argue that $\a=0$ is, in fact, the generic asymptotics. 

\ssec{Adding global charge}

The result \eqr{Dtowera} immediately generalizes if we add a density of states for $\O_p$ with asymptotics $\pst(\D_p\gg1) \sim p^{{\tt x}-1}$, whereupon $D=d+1+{\tt x}+\a$. This happens if---to take one physically  relevant example---the integral tower $\O_p$ is charged under a global symmetry $G$.
 
For concreteness, take $G=\mf{so}(\mf{n})$, and the $\O_p$ to furnish rank-$p$ symmetric traceless irreducible representations of $\mf{so}(\mf{n})$. The dimension of this representation equals the total density of states $\pst(\D_p)$, that is, the total number of operators in the rank-$p$ irrep. This dimension is known to be
\e{}{{(\mf{n}+2p-2)(\mf{n}+p-3)!\o p!(\mf{n}-2)!}}
At $p\gg1$, this scales as $p^{\mf{n}-2}$. This asymptotic density, in turn, modifies $D$: with the asymptotic scaling \eqr{cppp}, one finds
\e{}{D = d+\mf{n}+\a}
Assuming linearity of $C_{\phi pp}$ ($\a=0$), one has
\e{Ddn}{D=d+\mf{n}}
These are the symmetry and dimensionality of an AdS$_{d+1}\times S^{\mf{n}-1}$ compactification. Indeed, maximally supersymmetric SCFTs in $d=2,3,4,6$ have $\mf{so}(\mf{n})$ R-symmetry for appropriate values of $\n$, and towers of superconformal primaries in symmetric traceless representations thereof. The extension to products $\mf{so}(\mf{n}_1)\times \mf{so}(\mf{n}_2)$, etc, is immediate.

More generally, suppose $\O_p$ furnish a sequence of irreps $\R_p$ of some global symmetry $G$, indexed by $p$. Parameterize the dimension of $\R_p$ in terms of $p \approx \D_{p\gg1}$, where
\e{}{\pst(\D_{p\gg1}) = \text{dim}(\R_{p\gg1})~.}
Again assuming that $\a=0$, as we will for the rest of this section, if
\e{dimgrowth}{\text{dim}(\R_{p\gg1})\sim  p^{\r_p}}
then 
\e{Dr}{D=d+2+\r_p}
Dimensions of irreps of Lie algebras may be computed by Weyl's dimension formula. For $\mf{su}(\mf{n})$, for example, 
\e{}{\text{dim}([ a_1\ldots a_{\mf{n}-1}]) = \prod_{k=1}^{\mf{n}-1}\prod_{i=1}^k \left({\sum_{j=i}^k a_j\o k-i+1}+1\right)}
where $a_i$ are Dynkin labels. $[p\,0\ldots 0\,p]$ irreps of $\mf{su}(\n)$ grow as \eqr{dimgrowth} with $\r_p = 2\mf{n}-3$. This is relevant for the case of the $\cN=6$ ABJM theory in the type IIA `t Hooft limit at large $\D_\gap\sim\l^{1/4}$ \c{abjm}, whose superconformal primaries of $\D_p={p/2}$ furnish $[p\,0\,p]$ irreps of $\mf{su}(4)_R$ with $p\geq 2$. The above formulas imply $\r_p= 5$ and hence $D=10$, which saturates the correct bulk dimension.

\sssec{SCFTs}
The previous arguments allow us to associate a ``characteristic dimension'', $D_\cN$, to SCFTs with $\cN$ supersymmetries at large $c$ and large $\D_{\rm gap}$, if we assume the existence of a tower of operators $\O_p$. Let us take $d=4$ for concreteness. We have already treated the $\cN=4$ case above; below we treat $\cN=2,1$ (the case $\cN=3$ is easily constrained likewise).

\bul At $\cN=2$, the R-symmetry is $\mf{su}(2)_R\times \mf{u}(1)_r$. Towers of $\O_p$ are ubiquitous in $\cN=2$ SCFT, namely, as Higgs or Coulumb branch operators. The former are present in any Lagrangian $\cN=2$ SCFT with hypermultiplets, and are common in class $\mathcal{S}$ theories; e.g. in class $\mathcal{S}$ theories of genus-zero, the Higgs branch moduli space $\mathbf{M}_{\rm H}$ has complex dimension \c{Shimizu:2017kzs}
\e{}{\text{dim}_{\mathbb{C}}\mathbf{M}_{\rm H} = 24(c-a)}
where typically $c-a\sim N^\#$ for some $\#>0$. Denoting quantum numbers as $[j;\bar j]_\D^{(R;r)}$ where $j,\bar j$ are Lorentz spins, $R$ is the $\mf{su}(2)_R$ Dynkin label, and $r$ is the $\mf{u}(1)_r$ charge, the superconformal primaries have (e.g. \c{cdi})
\es{}{\text{Higgs branch,}~~ \hat{\mathcal{B}}_R:&\quad [0;0]_{2R}^{(R;0)}\\
\text{Coulomb branch,}~~ \Ec_r:&\quad [0;0]_{r}^{(0;r)}}
Assume that $C_{\phi pp} \sim p$ at large $p$. Alternatively, take $\phi$ to be the $\D=3$ scalar in the stress-tensor multiplet. Then by \eqr{Dr}, a tower of Higgs branch operators grows two large extra dimensions, while a tower of Coulomb branch operators grows one. Assuming their existence, the characteristic dimension of $\cN=2$ SCFT is
\es{}{\text{Higgs branch,}~~ \hat{\mathcal{B}}_R:&\quad D_{\cN=2}\geq 7\\
\text{Coulomb branch,}~~ \Ec_r:&\quad D_{\cN=2}\geq 6}
The $D_\cN$ should be viewed as bounds, not equalities, if only because the growth of ``generic'' single-trace operators with $\D_\O\neq \D_p$ can add more dimensions. 

Note that for a $\cN=2$ SCFT with a Higgs branch and a point $p_*$ on moduli space where $\D_\gap\gg1$, we bound $D_{\cN=2}\geq 7$ everywhere on moduli space continuously connected to $p_*$: this is because our calculation relies only on $C_{\phi pp}$ asymptotics, and Higgs branch three-point correlators are not renormalized \c{Baggio:2012rr, Beem:2013sza}.

\bul At $\cN=1$, the R-symmetry is $\mf{u}(1)_r$. The story is similar: given a tower of $\O_r$ with $\D\propto |r|$ -- e.g. (anti-)chiral primaries with $\D={3\o 2}|r|$ -- we infer
\e{Dn1}{D_{\cN=1}\geq 6}
This confirms the folklore expectation. Note the central role of the $\O_r$, and of the linearity of $C_{\phi rr}$: without them, the $\mf{u}(1)_r$ does not necessarily generate an extra dimension. The reader may note that the classic constructions of AdS/CFT pairs with 4d $\cN=1$ SUSY have $D=10$. We discuss this further in Section \ref{sechhc}. 

\ssec{$1/\D_\gap$ corrections and AdS quartic vertices}

Let us assume we have another scale in the flat space amplitude, denoted by $\ell_s$. Now the amplitude depends on the dimensionless combination $\ell_s\sqrt{s}$ and its high-energy limit is no longer fixed by dimensional analysis. For example, at tree-level,
\e{}{{L^{d-1}\o c}A_{d+1}^{\rm tree}(s,t) \sim {(L\sqrt{s})^{d-1}\o c}f_{d+1}^{\rm tree}\left(\theta;\ell_s\sqrt{s}\right)}
The CFT data will depend on a corresponding extra scale $\D_\gap$,\footnote{This scale does not have to coincide with the gap to higher spin single-trace operators, although in several contexts it does. With this understanding, we will keep using the name $\D_\gap$.} with the correspondence $\ell_s\sqrt{s} \sim n/\D_\gap$ (cf. \eqr{Lsn}), in the flat space limit. The $n$-scaling of OPE data in $\b^\1_{n,\ell}$ will exhibit crossover between the regime \eqr{regime}, which counts large bulk dimensions $L_\M \gg \ell_s$, and the regime $\D_\gap\ll n\sim \D_\O\ll c$, which also counts ``small'' extra dimensions $L_\M\lesssim \ell_s$. 

From the point of view of the effective theory in AdS, the derivative expansion generates a tower of quartic vertices which contribute to the anomalous dimensions of double-trace operators at $\O(c^{-1})$. To warmup, we first consider a quartic vertex without derivatives, $\a\phi^4$. From the CFT perspective this corresponds to a 'truncated' solution to the crossing equations where only scalar double-trace operators acquire an anomalous dimension. In $d=2$ and $d=4$, this solution was constructed in \c{hpps}. The expression for general $\Delta_\phi$ and $d$ is quite cumbersome, but for our purposes we are interested in the large $n$ behaviour. For a $\a\phi^4$ coupling \c{katz},
\e{}{\left. \gamma_{n\gg 1,0}^{(1)} \right|_{\phi^4} \sim \alpha \, n^{d-3}}
We can also consider quartic vertices with derivatives, schematically of the form $\alpha_q \phi^2 \partial^q \phi^2$, with $q=4,6,\cdots$. The large $n$ behaviour is now \c{katz, hpps}
\e{}{\left. \gamma_{n\gg1,\ell}^{(1)} \right|_{\phi^2 \partial^q \phi^2}\sim \alpha_q \, n^{d-3+q}}
with allowed spins $\ell\leq q/2$. Note that consistency with (\ref{Alarges}) in the flat space limit implies that all vertices with $q>2$ should be suppressed by powers of $1/\D_\gap$. Furthermore, we can estimate the suppression of each vertex $\alpha_q$. Since the flat space amplitude at high-energy is a function of $n/\D_\gap$ (due to the identification $\ell_s\sqrt{s} \sim n/\D_\gap$) and since the contribution of the stress tensor appears with coefficient of order one, we estimate
\e{suppress}{\a_q \sim {1\o (\D_\gap)^{q-2}},~~~ q \geq 2}
in analogy with effective field theory in flat space.\foot{Since no unsuppressed vertex grows with a higher power of $n$ than the stress tensor contribution, it is straightforward to check that at 1-loop (upon squaring the anomalous dimension) the presence of vertices will not change our conclusions about $D$. Said in reverse, one can also motivate the suppression \eqr{suppress} by demanding that at 1-loop, the vertices don't give a contribution to $\b_{n,\ell}^\1$ that exceeds the scaling set by $D$, where $D$ is read off from the squared stress tensor contribution to $\b_{n,\ell}^\1$ at sufficiently high $\ell$ where the vertices don't contribute.}

We now derive a new constraint for the case of quartic vertices involving heavier fields using the 1-loop sum rule. Let us assume the existence of the tower ${\cal O}_p$. We first consider truncated solutions to the tree-level correlator $\langle \phi\phi {\cal O}_p {\cal O}_p \rangle$. These correspond to AdS vertices of the form $ \alpha(p) \phi^2 \phi_p^2$ and similar vertices with derivatives. For each finite $p$, it turns out the flat space limit at tree-level does not impose further constraints, besides the suppression discussed above for vertices with more than two derivatives. But at $p\gg1$, we can constrain $\a(p)$ as follows. Let us focus on the vertex without derivatives. The solution in spacetime can be written as
\begin{equation}
{\cal A}_{\phi \phi pp}(z,\bar z) = \frac{\alpha(p)}{\Gamma(p-1)} u^p \bar D_{p p \Delta_\phi \Delta_\phi} 
\end{equation}
where the $\bar D$-function is the famous AdS quartic integral, see e.g. \c{DHoker:1999kzh}. We can then compute the contribution of this vertex to the tree-level anomalous dimension $\langle \gamma_{n,0}^{(1)}(p) \rangle$, upon solving the mixing problem. In the limit \eqr{npregime} with $ p/n \equiv x$ fixed, we find
\begin{equation}
\left.\langle \gamma_{n,0}^{(1)}(p) \rangle\right|_{\phi^2 \phi_p^2} \sim \alpha(p) n^{d-3} f_{\Delta_\phi}(x)
\end{equation}
where the form of $f_{\Delta_\phi}(x)$ will not be important for us, but it can be found for specific cases.\footnote{For instance, for $\Delta_\phi=2$ we find $$ f_{2}(x)= \frac{\Gamma\left( 3-\frac{d}{2}\right)}{2^{d}\Gamma\left( \frac{d}{2}\right)}(x+1)^{{d-4\o 2}}(2x+1)^{\frac{d-2}{4}}.$$} More generally, if we consider a vertex with $q$ derivatives we obtain 
\e{}{\left.\langle \gamma_{n,0}^{(1)}(p) \rangle\right|_{\phi^2 \p^q\phi_p^2}\sim \alpha_q(p)\, n^{d-3+q}f^{(q)}_{\Delta_\phi}(x)~.}
Now, plugging this into \eqr{b1} and the 1-loop sum rule \eqr{main} yields constraints on $\a(p)$. Consider for instance the vertex without derivatives, which is not suppressed by $\D_\gap$, and let us assume $\alpha(p\gg 1) \sim p^\sigma$.  The square of the anomalous dimension will contribute to $\b_{n,\ell}^\1$ as 
\begin{equation}\label{1loopalpha}
\b_{n,\ell}^\1 \supset \sum_{p=2}^{n}\pst(p)\,\alpha(p)^2 n^{2d-6} \sim n^{2d-5+2\sigma+\mathbf{r}_p}
\end{equation}
where we have allowed for asymptotic degeneracies of the $\O_p$ as $\pst(p\gg1) \sim p^{\mathbf{r}_p}$. Now we have a constraint: the growth in \eqr{1loopalpha} must not exceed $n^{2d-1+\mathbf{r}_p}$, where we have used $D=d+2+\mathbf{r}_p$.\foot{That $D=d+2+\mathbf{r}_p$ can be seen from looking at $\b_{n,\ell>0}^\1$, where the quartic vertex vertex does not contribute, and using \eqr{Dr}.} This implies $\sigma \leq 2$. Put in a different way, if $\alpha(p)$ grows faster than $p^2$, then the vertex $\phi^2 \phi_p^2$ must be suppressed by the gap scale, which would be inconsistent with bulk effective field theory. Similar bounds arise for each non-suppressed vertex.

The case of ${\cal N}=4$ SYM is conceptually similar, but the scalings are different because of supersymmetry. The picture at tree-level for the correlator $\langle {\cal O}_2{\cal O}_2{\cal O}_2{\cal O}_2 \rangle$, where $\O_2=\O_{\bf 20'}$, is the following, see \c{Alday:2014tsa}. The contribution from the stress tensor, namely the supergravity result, scales again like $n^{d-1} = n^3$. Vertices again scale as $n^{d-3+q}$, but there is an overall eight-derivative operator acting on the vertices, so that $q=8+\hat q$, with $\hat q=0,4,\cdots$. It then follows that already the first vertex is suppressed. In this case $\Delta_{\rm gap} \sim \lambda^{1/4}$, which corresponds to the mass of the string states, and the first vertex is suppressed by $\Delta_{\rm gap}^6$, in agreement with (\ref{suppress}). At 1-loop, it is important to take into account the tower of operators ${\cal O}_p$. The supergravity contribution and the first vertex, in the limit $n \gg 1$ with $p \sim n$, is given by
\begin{equation}
\left. \langle \gamma^{(1)}_{n,\ell}(p)\rangle \right|_{\rm sugra} \sim n^5,~~~~~\left. \langle \gamma^{(1)}_{n,\ell}(p)\rangle \right|_V \sim \alpha(p) \,n^7 \sim n^{11},~~~\alpha(p) \sim p^4 
\end{equation}
Note that the correct scaling for $\alpha(p)$ follows from the requirement that at 1-loop the `square' of the anomalous dimension in the flat space limit depends on the combination $n/\Delta_{\rm gap}$.

\sec{Bounding holographic spectra}\label{sec3}

Having derived from CFT some features of AdS$\times \M$ compactifications and values of $D$ for known setups, let us turn the logic around, by using our derivation to rigorously constrain possible behaviors of planar strongly coupled CFTs and of string theory. That is, we can use 1-loop unitarity in the flat space limit to ``bootstrap'' spectral properties of AdS/CFT dual pairs. We will show only a sample of what is implied by 1-loop unitarity in the flat space limit, leaving an exhaustive analysis of permissible symmetries, irreps and dimensions $D$ for the future.

First we focus on the single-trace growth, \eqr{pstx} and \eqr{d1x}. Insisting that the bulk does not have more than the 10 large dimensions of critical superstring theory implies that
\e{stringbound}{\pst(\D\gg1) \lesssim \D^{8-d}}
A similar bound applies for CFTs with M-theory duals,
\e{Mbound}{\pst(\D\gg1) \lesssim \D^{9-d}}
Conversely, polynomial growth that violates \eqr{Mbound} would imply that the ground state of the bulk dual is a stable AdS vacuum of supercritical string theory (or, perhaps, some other high-dimensional theory of quantum gravity).\foot{Most work on supercritical string solutions takes advantage of the rolling dilaton to generate time-dependent cosmologies \c{Silverstein:2001xn, Aharony:2006ra, Hellerman:2006nx, Green:2007tr}. In what follows we will continue to explore the consequences of forbidding supercritical AdS vacua in the bulk.} 

Note that in $d=2$, these are stronger constraints than the HKS \c{hks} sparseness bound on the total density of states, $\rho_{\rm HKS}(\D\leq {c\o 12}) \lesssim \exp(2\pi\D)$. That condition was shown to imply various hallmarks of semiclassical 3d gravity. Large $\D_\gap$ is not required for the HKS analysis. If $\D_\gap$ is large -- as it would be in the elusive prototypical CFT$_2$ which would saturate modular bootstrap bounds at large $c$ \c{Hellerman:2009bu, Friedan:2013cba, Collier:2016cls, Afkhami-Jeddi:2019zci, Hartman:2019pcd}-- the single-trace density must not grow super-polynomially in $\D$ for $\D\ll \D_\gap$. (At higher energies $\D\gtrsim\D_\gap$, exponential behavior of the total density of states, a la Hagedorn, can set in \c{Haehl:2014yla, Belin:2014fna}.)

Note also that ${\tt x}\notin \Z$ would imply $D\notin \Z$; one might reasonably like to avoid this, in which case our formula implies that ${\tt x}\notin \Z$ should not be possible. 

Next we turn to the result \eqr{Dr}, which is a linear constraint on the dimension $D$ and asymptotics $\r_p$ of the irrep $\R_p \in G$. This leads us to the following observations:

\begin{itemize}
\item We can constrain what irreps may be furnished by towers $\O_p$ in planar large $\D_\gap$ CFT. For example, one might speculate whether a 3d $\cN=6$ CFT at large $\D_\gap$ could have a tower $\O_p$ in the $[p\,p\,p]$ of $\mf{su}(4)_R$; this irrep has $\r_p=6$, hence $D=11$, which would be incompatible with a string theory dual. 

\item Conversely, imposing a critical superstring bound $D\leq 10$ implies a constraint on the symmetry groups $G$ and their irreps $\R_p$ that can be realized by local primary operators in large $c$, large $\D_{\rm gap}$ CFTs: for CFTs with string duals, only $$\r_p \leq 8-d$$ are consistent (similarly for M-theory). For instance, a CFT$_d$ with a critical string dual cannot have a tower of $\O_p$ in $\mf{su}(4)$ irreps $[p^2\,0\,p^2]$ in {\it any} $d$, because this irrep has $\r_p=10$. This proves one case of a result familiar from superconformal representation theory (though no SUSY was assumed): BPS operator dimensions are linear in charge. 

Indeed, the above result touches upon an inherent feature of holography: in KK reduction on AdS$\times \M$ in supergravity, $(mL_\M)^2$ equals an eigenvalue of a quadratic Laplace-type operator. This is equal, up to a $\D$-independent shift, to the quadratic Casimir $C_2$ of a highest-weight irrep of the isometry group $G_\M$; in the Dynkin basis,
\e{casimir}{C_2(\lbrace a_i\rbrace)\sim \sum_{i,j=1}^{\text{rank}(G_\M)}(a_i +2\rho_i) \,g^{ij}_\M\, a_j}
where $g^{ij}_\M$ is the inverse Cartan matrix of $G_\M$ and $\vec\rho$ is the Weyl vector. At large mass, 
\e{}{\D \sim mL_\M}
Therefore, the leading order $a_i\gg1$ asymptotics are linear in $\D$ for at least one index $a_{\hat i}$, and no faster than linear for the rest: 
\e{lineardynkin}{a_{\hat i} \sim \D~, \quad a_{j\neq \hat i} \lesssim \D}
Like $a=c$ \c{Henningson:1998gx, cemz}, this should be viewed as an emergent behavior of large $c$ CFT at large $\D_\gap$: it follows straightforwardly from properties of a two-derivative bulk, but from what does it follow in CFT? It would seem an interesting challenge to prove this in general, not just in certain examples.

\item If we {\it impose} that the CFT has a {\it critical} string or M-theory dual, we can infer the existence, or lack thereof, of operators of ``generic'' conformal dimension. Suppose that we insist upon $D=10$. Integral towers $\O_p$ can only generate this if $\r_p=8-p$, but this may not be compatible with a given global symmetry $G$ and irrep $\R_p$. In such a case, generic operators must make up the difference. e.g. take 
$G=\mf{so}(3)$ and $\R_p=p$. Then $\r_p=1$, so $D=d+3$. To get to $D=10$, one needs to add generic operators with asymptotic density $\pst(\D\gg1) \sim \D^{7-d}$.

\item This analysis also makes clear that large flavor symmetry groups $G_F$ with integral towers of local operators of increasing charge cannot be geometrized at the AdS scale: that is, charge-carriers of asymptotically large dimension $\D_p$ cannot have large charge $q_p$ -- more precisely, large $\mathbf{r}_p$ -- in the $N_F\rar\i$ limit. Towers of {\it fixed} charge can appear in AdS$\times \M$ compactifications as ``isolated'' vector multiplets of $G_F$ in the $D$-dimensional theory. Upon reduction on $\M$, they yield towers of $\O_p$ in the adjoint of $G_F$.\foot{Conclusions about the non-geometrization of large flavor groups do not hold at finite $\D_\gap$, where charged operators can be dual to non-decoupled string states.}

\end{itemize}

\sec{OPE universality at the string scale}\label{sec4}
 
Let us recap one of our previous results. Suppose a large $c$, large $\D_\gap$ CFT contains an integral tower of single-trace operators $\O_p$ with dimensions $\D_p=\D_\phi+{p-2}$ with $p=2,3,\ldots$ for some operator $\phi \equiv \O_2$. Parameterizing the OPE asymptotics as
\e{pasym}{C_{\phi pp}\big|_{p\gg1} \sim {p^{1+{\a\o 4}}\o \sqrt{c}}~,~~\text{where}~1\ll p \ll \D_\gap~,}
then $D=d+2+\a$. In this section we argue for a more general, and stringent, property of OPE asymptotics in CFT.
\vs
\begin{quotation} \noindent {\it {\bf Stringy OPE universality:} Consider a large $c$, large $\D_\gap$ CFT containing scalar single-trace operators $\phi$ and $\O_p$, with respective conformal dimensions $\D_\phi$ and $\D_p\propto p$, where}
\end{quotation}
\e{linearregime}{\D_\phi \ll \D_\gap\quad \text{and} \quad \D_\phi \ll p\ll c^{\#> 0}~,}
\begin{quotation} \noindent {\it Then the normalized planar OPE coefficient $C_{\phi pp}$ has linear asymptotics in $p$,}
\end{quotation}
\e{linear}{C_{\phi pp}\big|_{p\gg1} \sim {p \o\sqrt{c}}f(\D_\phi)}
\begin{quotation} \noindent {\it for some function $f(\D_\phi)$. If $\O_p$ has spin, then the leading scaling among all tensor structures obeys \eqr{linear}.}
\end{quotation}
\vs
\noindent ``Normalized'' refers to the norms of $\phi$ and $\O_p$. The $\O_p$ can be heavy KK modes with $\D_p\ll \D_\gap$; ``string-scale'' operators with $\D_p \sim \D_\gap$ (hence the title of the section); or semiclassical operators with $\D_\gap \ll \D_p \ll c^{\#>0}$. Note that $\D_p$ may be a function of global charges $\lbrace Q_i\rbrace$ that are held fixed in the limit, i.e. $\D_p \sim p f(Q_i)$. In such cases, the function $f(\D_\phi)$ in \eqr{linear} is also a function of these charges. The linearity property is also meant to apply to $C_{\phi p p'}$ where $|\D_{p'}- \D_{p}|/\D_p \ll 1$, but we continue to refer to $C_{\phi pp}$ for simplicity. 

The OPE coefficient $C_{\phi pp}$ at $p\gg1$ is what one might call a ``planar heavy-heavy-light'' three-point function, defined in a regime where perturbative large $c$ factorization applies to correlators.\foot{Heavy operators in 2d CFT with $1\ll\D\ll c$ are sometimes called ``hefty'' operators \c{dyer}.} This is {\it not} the same kind of heavy-heavy-light OPE coefficient featured in discussions of the Eigenstate Thermalization Hypothesis \c{Lashkari:2016vgj} (where $\D_{\rm heavy} \gg $ all other parameters), of CFT$_2$ \c{km} (where $\D_{\rm heavy} \gtrsim c$, dual to a conical defect or black hole microstate), or of the large charge expansion \c{Hellerman:2015nra} (where $\D_p \propto Q^{d\o d-1}\gg $ all other parameters, and $\O_p$ has charge $Q$ under a global symmetry). Indeed, in all of these cases the scaling is nonlinear. We are answering a different question, relevant to planar CFTs.

There are various arguments for \eqr{linear}:

\begin{boenumerate}

\item ``Naturalness'' of the emergence of large extra dimensions implies linearity, as discussed around \eqr{cppp}. This is the case whether the $\O_p$ are charged (see \eqr{Ddn} and \eqr{Dr}) or uncharged (see \eqr{Dtower}). 

\item Consider a toy model in which the integrally-spaced dimensions are deformed: $\D_{p\gg1}(\eps) \approx p(1+\eps)$. For finite $\eps$, they contribute to \eqr{b1} as ``generic'' single-trace operators, with $D$ determined by the asymptotic growth of $\pst(\D_p)$. If one demands that $D$ remains the same as $\eps\rar0$, linearity follows, by matching the scaling of \eqr{c2gap} and \eqr{gammap}. In other words, the first term of \eqr{b1} should be a limit of the second term of \eqr{b1}. 

\item Linearity is exact in certain cases: $C_{\phi pp}\propto \D_p$ when $\phi=\mathcal{L}$ or $T_{\mu\nu}$ or, in an SCFT, any other operator in their supermultiplets. On general grounds, the asymptotics should only depend on $\D_\phi$ in the overall coefficient: from the point of view of the heavy operator, the coupling to $\phi$ acts as a small perturbation. Moreover, the linearity does not depend in any way on the asymptotic spacing of the tower of $\O_p$. Together, these suggest linearity in general.

\item In Appendix \ref{appcppp}, we give a suggestive alternative argument. $C_{\phi pp}$ is dual to a three-point function of marginal vertex operators in an AdS$\times \M$ worldsheet CFT. At leading order in $1/\D_\gap$ perturbation theory, the worldsheet CFT, and the anomalous dimensions of vertex operators, effectively factorize. In particular, vertex operators dual to heavy operators have large but opposite anomalous dimensions in the AdS and $\M$ sigma models. This allows us to use known facts about heavy-heavy-light three-point functions in 2d CFT at {\it finite} $c$ \c{km, datta}. In Appendix \ref{appcppp}, we show how to reproduce linear scaling under certain assumptions.

\end{boenumerate}

\noindent We also find copious evidence for \eqr{linear} from existing calculations in planar CFT:

\begin{boenumerate}
\item In maximally supersymmetric CFTs in $d=2,3,4,6$, this holds for three-point functions of chiral primaries \c{Lee:1998bxa, Bastianelli:1999en,Lunin:2001pw}. This is simple to see for $d=3,4,6$. For $d=2$, the computation was performed in \c{Lunin:2001pw} at the symmetric orbifold point in the D1-D5 moduli space, but is valid at $\l\gg1$ thanks to a non-renormalization theorem \c{Baggio:2012rr}. Their result is rather complicated, but has linear scaling, as we show in Appendix \ref{appLM}. 

\item Extremal chiral primary three-point functions in the `t Hooft limit of the ABJM theory also obey linearity \c{Hirano:2012vz, bk}. 

\item In planar $\cN=4$ SYM at strong coupling, there are many examples involving a diverse set of operators: protected and unprotected; twist-2 and large twist; spinning and not spinning. In what follows we denote Lorentz spin as $S$ and R-charge as $J$:

{\addtolength{\leftskip}{10 mm}\addtolength{\rightskip}{10 mm}
\begin{itemize}
\item When $\phi = \mathcal{L}$ or a superpartner thereof and $\O_p$ is any heavy operator where $\D_p$ obeys power-law scaling in $\l$ \c{costa,zar}. In this case, linearity follows from $\la \mathcal{L}\, \O_p\O_p \ra \propto  \l{\p \D_p \o \p\l}$. This was confirmed in explicit computations (e.g. \c{costa, zar, mp, roiban}).

\item When $\phi$ is a light chiral primary and $\O_p$ is a ``short'' string state. In particular, $C_{\phi pp}$ was computed with $\O_p$ as a Konishi-like operator at the first massive level ($S=0, J\sim \O(1)$, and $\D_p\sim \l^{1/4}$) in \c{bmp}, and as a twist-2 massive Regge state of nonzero spin ($S\sim \O(1), J\sim \O(1)$ and $\D_p\sim \sqrt{2(S-2)}\l^{1/4}$) in \c{mp}.\foot{If $\O_p$ is a (non-conserved) symmetric traceless tensor of spin-$S$, then $C_{\phi pp}$ has $S+1$ independent tensor structures \c{cppr}. In the computation of \c{mp} and the notation of \c{cppr, mp}, the structure that dominates at large $\l$ is $H_{23}^S$. (See eqs. (4.36) and (4.55) of \c{mp}.)}. All results are linear in $p\equiv\l^{1/4}$. Note that these heavy operators are {\it unprotected} and have order one R-charge, so this result does not follow from symmetry, or KK considerations, alone. 

\item When $\phi$ is a light operator and $\O_p$ is a ``semiclassical'' state with $\D_p\sim\sqrt{\l}$. This case is amenable to bulk worldsheet techniques \c{costa, zar}. This includes BMN operators \c{zar}, both chiral and non-chiral primaries with large charge $J\sim\sqrt{\l}$; spinning strings in $S^5$ \c{zar}; spinning strings in AdS$_5$ \c{roiban}; and strings with spin in both AdS$_5$ and $S^5$  \c{roiban}. All results are linear in $p\equiv\sqrt{\l}$. 

\end{itemize}}
\end{boenumerate}

There are some cases where the scaling of $C_{\phi pp}$ is sub-linear, that is, $f(\D_\phi)=0$. Aside from cases where selection rules forbid nonzero $C_{\phi pp}$, this can happen when $C_{\phi pp}$ is  controlled by symmetry: for example, if $\phi$ is in a flavor current supermultiplet such that $C_{\phi pp}\propto q_p$, and the $\O_p$ do {\it not} have charge increasing with $\D_p$.\foot{One such case is if $\phi$ is the moment map operator $\hat \cB_1$ in a 4d $\cN=2$ SCFT, and the $\O_p$ are all in the adjoint of the flavor symmetry $G_F$. This happens holographically when the $\O_p$ are all KK modes of a $D$-dimensional $G_F$ vector multiplet, as mentioned in the previous subsection. We thank Ying Hsuan-Lin and David Meltzer for discussions on this.} Such special cases aside, we believe the linear scaling to be generic. 

It would be worth understanding the widest regime in which linearity holds. To this end, we note that in the `t Hooft limit of ABJM and in $\cN=4$ SYM, three-point functions where $\phi$ is a light chiral primary and $\O_p$ is a giant graviton {\it still} obey \eqr{linear}, despite the giant graviton having $\D\sim N$ \c{Hirano:2012vz}. In particular, the results of \c{Hirano:2012vz} take the limit $\D/N$ fixed; subsequently taking the ratio to be small, one finds linearity in $\D$. Finally, though our result is meant to apply only at large $\D_\gap$, in 4d $\cN=2$ $\mf{su}(N)$ SQCD in the large $N$ Veneziano limit, the $tt^*$ analysis of  \c{Baggio:2016skg} shows that three-point functions of chiral primaries do obey linearity to first non-trivial order in $\l\ll1$ (see eq. 4.20 therein).

\ssec{Asymptotic linearity of planar non-BPS spectra}\label{nonbps}

\eqr{linear} has an implication for asymptotic scaling of non-BPS spectra of planar SCFTs. 

Take $\phi$ to be a scalar component of an R-current supermultiplet, and take $\O_p$ to be a bottom component of a long supermultiplet, for which $C_{\phi pp}$ is proportional to the R-charge of $\O_p$. The unitarity bound for long supermultiplets states that $\D_p> \sum_i b_i R_p^{(i)} + b$, where the sum runs over the Dynkin indices of the R-symmetry irrep, for some constants $b$ and $\lbrace b_i\rbrace$ that depend on the spacetime dimension and number of supersymmetries. Suppose there exists a family of such long multiplets $\O_p$ of increasing charge, with fixed $p$-dependence of $R^{(i)}_p$. For simplicity of presentation, let us turn on just a single index $R_p\equiv R_p^{(1)}$. At large charge $1\ll R_p\ll c^{\#>0}$, unitarity requires only that
\e{}{\D_p \,\gtrsim \,\widehat b_1 R_p+\ldots~,~~\text{with}~~\widehat b_1  \geq b_1} 
In particular, the scaling in $R_p$ must be at least linear, with bounded coefficients in the linear case. On the other hand, {\it super}-linear scaling would violate \eqr{linear}. Therefore, to leading order in large $p$, \eqr{linear} implies that the long operators $\O_p$ must have dimensions {\it linear} in the charge, with coefficient $\widehat b_1  \geq b_1$. 

This behavior is familiar in some planar CFTs, e.g. for BMN operators in $\cN=4$ SYM which furnish $[0J0]$ irreps of $\mf{su}(4)_R$ and have $\D\sim J$ \c{bmn}. In the event that, in fact, $\widehat b_1  = b_1$, the long operators are hitting the unitarity bound asymptotically. In this sense, these non-BPS operators are becoming ``asymptotically free'' at large charge. Extrapolation of linearity to infinite charge, however, is cut off by the central charge $c$.\foot{The large charge analysis of \c{Hellerman:2017veg} found that in {\it non-planar} 3d $\cN=2$ SCFTs with a one-complex-dimensional moduli space, the conformal dimension of the lightest long supermultiplet with large R-charge obeys asymptotic linearity, with calculable subleading corrections. That echoes what we have found for planar SCFTs. We thank Clay C\'ordova for pointing this out. It would be interesting to classify the allowed crossovers between the regime \eqr{linear} and the regime $c \ll \D_p$, and how this depends on the existence of moduli. }

\sec{On holographic hierarchies}\label{sechhc}
This section is more speculative than, and somewhat detached from, the rest of the paper. 

Building on insights from the 1-loop sum rule \eqr{main}, our goal is to formulate a conjecture about the AdS/CFT landscape with large extra dimensions and to examine the ingredients involved. A good starting point is the condition explored by \c{ps, joetalk}:
\vs
\centerline{{\it Strong coupling + no R-symmetry  $\Rightarrow$ Local AdS$_{d+1}$ dual}}
\vs

\noindent ``Local AdS$_{d+1}$ dual'' means $D=d+1$. The spirit of the conjecture is likely correct. Our results allow us to hone this a bit. 

First, it is important to specify what abstract notion of ``strong coupling'' is required. A necessary condition for what is usually intended is large $\D_\gap$ \c{hpps}, but often some sense of ``sparseness''  of low-spin operators is left implicit. Our sum rule shows that large $\D_\gap$ {\it and} sub-polynomial growth of the $\ell\leq 2$ single-trace density, $\rho_{\rm ST}(\D)$, imply AdS$_{d+1}$ locality. This refinement is paramount: all fully controlled examples of holographic CFTs with large $\D_\gap$ also have large extra dimensions. 

So, with large $\D_\gap$ as a proxy for strong coupling, the veracity of the AdS locality claim boils down to whether ``no R-symmetry'' implies sub-polynomial single-trace degeneracy. 

To that end, we note that there are many examples in which there {\it is} R-symmetry, but the value of $D>d+1$ is due {\it not} to the protected operators, but to the {\it unprotected} ones. In other words, there is stronger polynomial growth of $\pst(\D)$ than is required by R-symmetry, or other accidental symmetries, alone. This is visible in the canonical constructions of AdS/CFT pairs with 4d $\cN=1$ SUSY, the infinite class of quiver gauge theories dual to AdS$_5\times SE_5$, where $SE_5$ is a Sasaki-Einstein manifold. These are Freund-Rubin solutions of type IIB supergravity, with $L_{SE_5} \sim L$. The bound \eqr{Dn1} makes clear that the existence of these critical string backgrounds is due not to the R-symmetry or to its towers of (anti-)chiral primaries {\it per se}, but to the Weyl's law scaling of {\it unprotected} single-trace operators. The KK spectrum of $SE_5 = T^{1,1}$, dual to the conifold CFT \c{Klebanov:1998hh}, was nicely studied in \c{gubser, ferrara1, ferrara2}. Towers of unprotected operators with finite irrational dimensions, despite the infinite coupling, are strikingly visible.\foot{$T^{1,1}$ has $\mf{u}(1)_r\times\mf{su}(2) \times \mf{su}(2)$ isometry, so one might suspect that our arguments secretly imply $D=10$ (10=5+2+2+1). But this is misleading. A more representative case familiar to holographers is that of $L^{pqr}$, which possesses only $\mf{u}(1)^3$ isometry ($5+3\neq 10$) \c{Cvetic:2005ft}. Infinite families of $SE_5$ manifolds have just the requisite $\mf{u}(1)$ Reeb vector \c{Sparks:2010sn}, though their stability as type IIB backgrounds is not guaranteed.} Likewise for KK reduction on AdS$_4\times M^{111}$ \c{fabbri}. 

A last minor point is that there are non-SUSY backgrounds with $D>d+1$ that are known to be at least perturbatively stable, including $T^{p,p}$, $M^{pqr}$ and $Q^{pqr}$ for certain non-SUSY values of $(p,q,r)$ \c{page1, page2, yasuda}. They may, of course, be non-perturbatively unstable.

Given this, one might suggest that {\it(local)  isometry}, not SUSY or R-symmetry, is the key concept. As in the known examples like $L^{pqr}$, extra isometries play an important indirect role in determining $D$, by protecting long multiplets from infinite mass renormalization at strong coupling. The many extra operators are conceivably stabilized by the presence of isometry at the AdS scale. Additionally, in SUSY-breaking double-trace flows at large $N$ \c{Dong:2015gya, banksov, Giombi:2017mxl} isometries guard against certain types of perturbative and non-perturbative instabilities, helping to furnish possible counterexamples to \c{ov, fk}. Taking this idea seriously leads to the following version of the conjecture of \c{ps, joetalk}: 
\vs
\vs
\centerline{{\bf Holographic Hierarchy Conjecture (HHC):} }
\vs
\centerline{\it Large $\D_\gap$ + no global symmetries $~ \Rightarrow~$ Local AdS$_{d+1}$ dual}
\vs
\noindent The HHC is quite similar to the conjecture of \c{ps, joetalk}. The HHC is weaker in the sense that it allows putative CFTs without R-symmetry, but with other global symmetries, to have large extra dimensions; this includes the solutions of \c{Dong:2015gya, banksov, Giombi:2017mxl}, which have no R-symmetry and may plausibly survive a stability analysis. On the other hand, if SUSY-breaking backgrounds at large volume turn out to be unstable \c{fk, ov}, then the HHC is essentially identical to the conjecture of \c{ps, joetalk} (upon replacing ``strong coupling'' by ``large $\D_\gap$''). Note that the HHC applies to sequences of CFTs with parametrically large $\D_\gap$, not to putative isolated CFTs with large but finite $\D_\gap$ and sporadic spectra. Ultimately, what one is looking for is a physical condition on such sequences that guarantees sub-polynomial growth of single-trace degeneracies; whether global or R-symmetry does so, or whether symmetry and degeneracy are in fact independent concepts that happen to share a place under the lamppost\foot{For example, the existence of SUSY solutions with large extra dimensions and fluxes which break all metric isometries besides a $\mf{u}(1)_R$ may be reasonably viewed as typical representatives of large-volume solutions. With respect to the HHC, the question then shifts to what becomes of such solutions when a $\mf{u}(1)_R$ is not present. We thank Ofer Aharony for his comments.}, is yet to be established.

The HHC has two corollaries which can be phrased purely in terms of CFT and gravity:
\vs
\begin{quotation} \noindent {\it {\bf Corollary 1 (CFT):} Large $\D_\gap$ + no global symmetries $~ \Rightarrow~$ Sub-polynomial growth of $\rho_{\rm ST}(\D)$ for $1\ll \D \ll \D_\gap$.}
\end{quotation}

\begin{quotation} \noindent {\it {\bf Corollary 2 (Gravity):} In two-derivative gravity coupled to low-spin ($\ell \leq 2$) matter, there are no stable AdS $\times \M$ solutions where $\M$ is a positively curved, compact Einstein manifold without local isometries.}
\end{quotation}

\noindent Corollary 1 follows from the HHC and our sum rule for dDisc$(\cA^\1)$, as discussed above. Corollary 2 follows from the HHC and the Freund-Rubin solution of (super)gravity. Intriguingly, Corollary 2 benefits from circumstantial evidence from differential geometry (e.g. \c{besse, anderson, yang}): positively curved, compact Einstein manifolds $\M$ with dim$(\M)>3$ and no local isometries are not known to exist! If they do exist, the conjecture is that they are unstable solutions of the $D$-dimensional theory.

Does the arrow run the other way? Even for R-symmetries, we know of no solid argument that there must be infinite towers of operators that populate the BPS charge lattice, and hence that R-symmetries must be geometrized at the AdS scale.\foot{This does not follow from the ``Completeness Conjecture'' or its modern avatar, ``Conjecture 2'' of \c{ho1, ho2}. These conjectures only require that, for some global symmetry $G$ (suitably defined), the CFT contains local operators which realize all finite-dimensional irreps of $G$; but the conformal dimensions of these operators are not fixed, so they can be heavy. One might ponder some form of a ``Superconformal Completeness Conjecture'' -- that there exist local operators which realize all finite-dimensional irreps of superconformal symmetry. But even accounting for the continuum of $\D$ allowed in long multiplets, the most literal form of this is incorrect, as many protected conformal multiplets are not realized in actual SCFTs. Perhaps the idea can be salvaged under certain conditions.} This is a compelling open question.

\sec{Future directions}

An obvious challenge for string/M-theorists and holographers is to construct bona fide vacua of small curvature. Even rigorous {\it bounds} on distributions of parametric hierarchies, as opposed to explicit solutions, would be quite welcome. We have not pursued that here, but we expect our 1-loop bootstrap perspective to be a useful tool in this endeavor. 

As a starting point, can we at least identify certain candidate CFTs with large $c$ and large $\D_\gap$ that have $D<10$? Strategically speaking, focusing on $d+1<D<10$ seems simpler than requiring a hierarchy for all extra dimensions. One suggestion is the class of 4d Lagrangian SCFTs studied in \c{Buchel:2008vz}. These are superconfomal gauge theories with $a=c$ to leading order in $1/N$. Their analysis included models that, unlike the Veneziano limit of $\cN=2$ $\mf{su}(N)$ SQCD which possesses a large number of {\it fundamental} fields $N_f = 2N$, achieve conformality with a {\it finite} number of  {\it non-fundamental} fields. The presence of an exactly marginal gauge coupling raises the question of whether, at strong coupling, any of these theories has large $\D_\gap$ and $D<10$. Entry (d) of the table on p.19 of \c{Buchel:2008vz} is particularly appealing:\foot{This is the theory in the fifth bullet point of Sec. 3.5.5 of \c{Bhardwaj:2013qia}.} it has $\cN=2$ SUSY, and $a-c\sim \O(1)$ instead of $\O(N)$ (or, worse yet, $\O(N^2)$, as in the Veneziano limit). The latter property suggests a closed-string dual. Thus, we have a possible candidate for a large $N$ CFT with $\cN=2$ SUSY, large $\D_\gap$, and a $D<10$ closed-string dual. It would be interesting to look into this more closely; a good starting point would be to study the BPS spectrum of this and other related models. 

Going beyond probing only {\it large} extra dimensions would obviously be worthwhile. Our formula already contains some hints of small extra dimensions. The function $f^\1(\ell)$ in \eqr{main} contains the angular structure of the $D$-dimensional S-matrix. This indirectly encodes the structure of small internal dimensions in the bulk (if any), because the S-matrix depends on the internal manifold on which string/M-theory is compactified. For instance, loop-level amplitudes of type IIB strings compactified on $K3$ vs. $T^4$ will depend on the number of 6d tensor multiplets. What can bootstrap techniques say about what type of functions $f^\1(\ell)$ are allowed to appear?

Finally, having found a dictionary for determining $D$ from planar OPE data, we only scratched the surface of actually bootstrapping the landscape. We have not, for instance, addressed the question of whether, in the presence of R-symmetry, infinite towers of light BPS operators are required for UV consistency. 

\section*{Acknowledgements} 
We thank Ofer Aharony, Chris Beem, David Berenstein, Agnese Bissi, Scott Collier, Frank Coronado, Xi Dong, Sergei Gukov, Ben Heidenreich,  Ying Hsuan-Lin, Shamit Kachru, Shota Komatsu, Dalimil Maz\'a\v c, David Meltzer, Yi Pang, Leonardo Rastelli, David Simmons-Duffin and Arkady Tseytlin for helpful discussions. This project has received funding from the European Research Council (ERC) under the European Union's Horizon 2020 research and innovation programme (grant agreement No 787185). We thank CERN during the workshop ``Advances in Quantum Field Theory,'' and the International Institute of Physics, Natal during the workshop ``Nonperturbative Methods for Conformal Theories,'' for stimulating environments. EP thanks the University of Michigan and KITP for hospitality during the course of this work. EP is supported by Simons Foundation grant 488657 and the U.S. Department of Energy, Office of Science, Office of High Energy Physics, under Award Number DE-SC0011632.

\appendix

\sec{Supplementary details at 1-loop}\label{appa}

\ssec{$C_{\phi\phi[\O\O]_{n,\ell}}$ at large $\D_\gap$}\label{appCa}

Our goal is to show that \eqr{c2gap} holds. This can be extracted from the four-point function $\la \phi\phi\O\O\ra$ at $\O(1/c)$ in the s-channel, $\phi\phi \rar \O\O$, which contains $[\O\O]_{n,\ell}$ exchange:
\e{}{\la \phi(0)\phi(z,\zb)\O(1)\O(\infty)\ra \supset (z\zb)^{-{\D_\phi+\D_{\O}\o 2}}\sum_{n,\ell}C_{\phi\phi[\O\O]_{n,\ell}} \sqrt{a^{(0)}_{[\O\O]_{n,\ell}}} G_{{[\O\O]_{n,\ell}}}(z,\zb)}
where $a^{(0)}_{[\O\O]_{n,\ell}}$ is the squared OPE coefficient of the MFT for $\O$. Universality at large $\D_{\rm gap}$ in the $1/c$ expansion implies that the $n\gg1$ scaling of $C_{\phi\phi[\O\O]_{n,\ell}} \sqrt{a^{(0)}_{[\O\O]_{n,\ell}}} $ is the same as that of $a^{(1)}_{[\O\O]_{n,\ell}}$, the $\O(1/c)$ correction to $a^{(0)}_{[\O\O]_{n,\ell}}$. This obeys 
\e{}{a^{(1)}_{[\O\O]_{n,\ell}} = {1\o 2} \p_n(a^{(0)}_{[\O\O]_{n,\ell}}\g^{(1)}_{[\O\O]_{n,\ell}}) + a^{(0)}_{[\O\O]_{n,\ell}}\hat a^{(1)}_{[\O\O]_{n,\ell}}}
where, as argued in \c{Alday:2017gde}, 
\e{}{\hat a^{(1)}_{[\O\O]_{n,\ell}}\big|_{n\gg1} \sim n^{-2\t_*}}
where $\t_*$ is the twist of the lowest-twist operator in the $\O\times \O$ OPE. Hence $\hat a_{[\O\O]_{n,\ell}}$ is negligible. Since $a^{(0)}_{[\O\O]_{n,\ell}}\sim 4^{-2n}\times $(power law), we have $a^{(1)}_{[\O\O]_{n,\ell}} \sim a^{(0)}_{[\O\O]_{n,\ell}}\g^{(1)}_{[\O\O]_{n,\ell}}$ at $n\gg1$. \eqr{c2gap} follows. 

\ssec{Computing $\la \g_{n,\ell}(p)\ra$}\label{appgamma}
Consider a CFT with a protected single trace scalar operator $\phi$ together with a tower of protected scalar operators ${\cal O}_p$, such that $\Delta_p  = \Delta_\phi+\Z_{>0}$. Let us take for definiteness $\Delta_\phi=2$ and $\Delta_p=p>2$. In this case, to order $1/c$, the following double-trace operators mix:
\begin{equation}
[\phi\phi]_n,~[{\cal O}_3{\cal O}_3]_{n-1},\cdots, [{\cal O}_{n+2}{\cal O}_{n+2}]_0
\end{equation}
where we have suppressed the spin index. The conformal partial wave decomposition at (approximate) twist $2\Delta_\phi+2n$ will then have a sum over  $n+1$ species. The mixing problem can be solved by considering the correlators $\langle \phi \phi \phi \phi \rangle, \cdots \langle \phi \phi {\cal O}_{n+2}{\cal O}_{n+2} \rangle$ (see e.g. \c{Aprile:2017xsp,Alday:2017xua}). It can be shown that the average squared anomalous dimensions appearing in the 1-loop correlator $\langle \phi \phi \phi \phi \rangle$ are given by
\begin{equation}
\langle \big( \gamma_{n,\ell}^{(1)}\big)^2 \rangle = \sum_{p=2}^{n+2} \langle \gamma_{n,\ell}^{(1)}(p) \rangle^2
\end{equation}
where $\langle \gamma_{n,\ell}^{(1)}(p) \rangle$ is the average anomalous dimension arising from the correlator $\langle \phi \phi {\cal O}_{p}{\cal O}_{p} \rangle$. Our strategy to compute $\langle \gamma_{n,\ell}^{(1)}(p) \rangle$ is as follows. Consider the correlator $\langle \phi \phi {\cal O}_{p}{\cal O}_{p} \rangle$. In the t-channel the operator ${\cal O}_{p}$ itself is exchanged: $\phi \times{\cal O}_p \to {\cal O}_p \to \phi\times {\cal O}_p$. This leads to the following contribution:
\begin{equation}
{\cal A}_{\phi \phi pp}(z,\bar z) \supset C_{\phi p p}^2 \frac{(z \bar z)^{\Delta_\phi}}{((1-z)(1-\bar z))^{\Delta_\phi/2}} g_{\Delta_p}(1-\bar z,1-z)
\end{equation}
where the conformal block $g_{\Delta_p}(1-\bar z,1-z)$ has been defined such that the small $1-\bar z$ dependence is explicitly shown and $C_{\phi p p}$ is the OPE coefficient $\langle \phi {\cal O}_p {\cal O}_p \rangle$. This contribution produces a singularity/double discontinuity around $\bar z=1$, which for the case $\Delta_\phi=2$ is a single pole. The residue of this single pole contains a piece proportional to $\log z$ which is independent of the number of spacetime dimensions,
\begin{equation}
{\cal A}_{\phi \phi pp}(z,\bar z) \supset C_{\phi p p}^2 \frac{1}{1-\bar z} (1-p) \left(\frac{z}{z-1} \right)^p \log z + \cdots
\end{equation}
where we are disregarding pieces not proportional to $\log z$ and which do not contribute to the double-discontinuity. This singularity can only be reproduced by the correct anomalous dimensions for double-trace operators propagating in the s-channel, with twist $2p+2m$, with $m=0,1,2,\cdots$. This anomalous dimension can be computed either by large spin perturbation theory or by the Lorentzian inversion formula. For fixed $m$ the computation can be done in general dimensions, but we were able to find a closed expression only for $d=2,4$. The final expressions are quite cumbersome, but they simplify in the large $n,p$ limit with $p/n=x$ fixed, where $2p+2m\approx 4+2n$ is the twist. In this limit we obtain 
\begin{eqnarray}
\langle \gamma_{n,\ell}^{(1)}(p) \rangle^2\Big|_{n\sim p\gg1} &\approx& C_{\phi p p}^4 \,n^2 x^4 \, \frac{(1-x)^{\ell+1}}{(1+\ell)^2(1+x)^{\ell+1}},~~~~d=4\\
\langle \gamma_{n,\ell}^{(1)}(p) \rangle^2\Big|_{n\sim p\gg1}   &\approx& 4 C_{\phi p p}^4 \, \frac{x^2}{n^2} \, \frac{(1-x)^{\ell}}{(1+x)^{\ell}},~~~~d=2
\end{eqnarray}
where $C_{\phi p p}$ is evaluated at $p\gg1$. Note that the scaling with $n$ is enhanced for the particular limit we are considering, with $p \sim n$, compared to the usual bulk-point limit $n\gg1$ with $p$ fixed. For instance, for finite $p$ and in $d=4$, one has $\gamma_{n\gg1,\ell}^{(1)} \sim 1/n$ \c{Alday:2017gde}. It is also possible to perform the computation in $d=6$, since in this case the conformal blocks are also known. Since $\Delta_\phi=2$ is an integer dimension below the unitarity bound for $d=6$, the result is actually divergent as $\Delta_\phi \to 2$, but this divergence does not affect the $n$ scaling.  It can be checked that $\langle \gamma_{n,\ell}^{(1)}(p) \rangle^2\Big|_{n\sim p\gg1} \approx C_{\phi p p}^4 \,n^6$ for $d=6$. The above results for $d=2,4,6$ suggest the following scaling for general $d$:
\begin{eqnarray}\label{gsqapp}
\langle \gamma_{n,\ell}^{(1)}(p) \rangle^2\Big|_{n\sim p\gg1}  &\approx& C_{\phi p p}^4\big|_{p\gg1}\, \,n^{2d-6}~.
\end{eqnarray}
This is the result quoted in the body of the paper. 

\sec{OPE universality: Further comments and computations}\label{applinear}

\ssec{Linearity of $C_{\phi pp}$ from the worldsheet}\label{appcppp}
The goal is to try to prove \eqr{linear} from a bulk worldsheet computation.\foot{We thank Shota Komatsu for helpful conversations.} Here we give a suggestive argument. We assume a $D$-dimensional AdS$\times \M$ solution of string theory dual to a large $c$, large $\D_\gap$ CFT. Consistent with the Holographic Hierarchy Conjecture of Section \ref{sechhc}, we assume that $\M$ has local isometries.
On the worldsheet lives a conformal sigma model, $\sigma_{\rm AdS\times\M}$, with coupling $1/\D_\gap$. (We are treating $\D_\gap$ as $\sim \l^{1/4}$.) We label the vertex operators dual to $\phi$ and $\O_p$ as
\e{}{V_L \leftrightarrow \phi~, \quad V_H \leftrightarrow \O_p}
We sometimes use the shorthand $L,H$. The boundary dimension of $\O_p$ is $\D_H$, which we take to be in the range $1 \ll \D_H \ll c^{\#>0}$. 

At leading non-trivial order in large $c$ and large $\D_\gap$, the AdS and $\M$ sigma models factorize, i.e. ``$\sigma_{\rm AdS\times\M} = \sigma_{\rm AdS}\times \sigma_\M$'': in particular, vertex operator dimensions obey
\e{}{\D_V = \D_V^{\rm AdS} + \D_V^\M}
and likewise, OPE coefficients factorize as
\e{}{C_{HHL} = C_{HHL}^{\rm AdS} C_{HHL}^{\M}}
We define these to be norm-invariant,
\e{}{C_{HHL} \equiv {\la V_HV_HV_L\ra\o \la V_HV_H\ra\sqrt{\la V_LV_L\ra}}}
Both $V_L$ and $V_H$ have total worldsheet dimension two,
\e{}{\D_i^{\rm AdS} + \D_i^\M = 2~,~~\text{where}~~ i=L,H~.}
Given an explicit ansatz for $V_L$ and $V_H$ in terms of worldsheet elementary fields, this in principle fixes the dimension of the boundary operator in the CFT \c{Polyakov:2001af}.\foot{See \c{Tseytlin:2003ac} and references thereto for concrete implementation of this idea. It has been demonstrated to work for certain classes of operators, but more generally its status is somewhat unclear.}

Now, we take $\D_H^\M$, the dimension of $V_H$ on $\sigma_\M$, to be proportional to $\D_H$, i.e. $$\D_H^\M \sim \D_H\gg1~.$$ This will be the case for heavy operators, namely, those with large classical dimension on $\M$, or with boundary dimension $\D\gg \D_\gap$ and hence large 1-loop dimension on $\M$.\foot{Since there is no general prescription for constructing explicit vertex operators on arbitrary $\sigma_\M$ (nor even on $\sigma_{{\rm AdS}_5\times S^5}$), we cannot be more explicit. In $\cN=4$ SYM, chiral operators or short string states with $\D\sim \l^{1/4}$ do not map to worldsheet vertex operators $V_H$ with $\D_H^{S^5}\gg1$. But semiclassical states, with $\D\sim \sqrt{\l} \sim \D_\gap^2$, do.} Then the OPE coefficient $C_{HHL}^{\M}$ is a heavy-heavy-light (HHL) three-point function in a 2d CFT at finite worldsheet central charge $c$. In 2d CFT, HHL leading asymptotics are fixed by modular covariance of torus one-point functions \c{km}. The original result of \c{km} applied to operators uncharged under any global symmetry, but was subsequently extended to include an affine $\mf{u}(1)_k \times \overline{\mf{u}(1)}_k$ current algebra \c{datta}, whereby torus one-point functions become weak Jacobi forms. $\sigma_\M$ has local currents due to the local isometries of $\M$. Taking $\O_L$ to be uncharged and $\O_H$ to be charged under an affine $\mf{u}(1)_k \times \overline{\mf{u}(1)}_k$ subalgebra, then for $\D_H\gg1$, the leading-order result is, up to an overall constant independent of heavy parameters,
\e{chargedhhl}{{C_{HHL}} \propto C_{\chi^\dag\chi L}\D_H^{\D_L/2} \exp\left(-\pi\sqrt{{4c\o 3}}\sqrt{\D_H}\left(1-\sqrt{1-{12\o c}\left(\D_\chi - \D_\chi^{\rm min}\right)}\right)\right)e^{-{2\pi i\o k}(q_\chi Q - \bar q_\chi \bar Q)}}
$\chi$ is defined to be the lightest charged operator with $C_{\chi^\dag\chi L}\neq 0$, and the unitarity bound is
\e{}{ \D_\O \geq \D_\O^{\rm min}\equiv{q_\O^2+\bar q_\O^2\o 2k}}
This is the OPE coefficient for an ``average'' individual heavy operator of dimension $\D_H$, which is what we want here. In writing the result of \c{datta} this way, we take $\O_H$ to be well above the unitarity bound, $\D_H-\D_H^{\min} \propto \D_H$. We draw attention to the power-law factor. 

Now we want to apply this to $\sigma_\M$. In the large-volume limit of large $\D_\gap$, $\sigma_\M$ is noncompact, with a mass gap $m\sim 1/\D_\gap$. The formula \eqr{chargedhhl} makes sense in a noncompact CFT because the density of states is weighted by the three-point functions $C_{\chi^\dag\chi L}$, which are nonzero by definition. In a noncompact CFT where, moreover, the charged continuum starts at the unitarity bound, \eqr{chargedhhl} reduces to (again dropping overall constants)
\e{}{{C_{HHL}}\sim C_{\chi_{\rm min}^\dag \chi_{\rm min} L}\D_H^{\D_L/2}}
Assuming this is the case for $\sigma_\M$ in the free limit, we arrive at
\e{HHLM}{C_{HHL}^\M \sim (\D_H)^{\D_L^\M/2}~,}
which is valid to leading order in $1/\D_\gap$. 

In order to to get the full result, we must combine this with the AdS side. Upon imposing marginality, $\D_H^{\rm AdS}$ is large and negative. This reflects the negative curvature of $\sigma_{\rm AdS}$. In the same spirit as other settings in which scaling dimensions are computed in $\sigma_{\rm AdS_{d+1}}$ by analytic continuation from $\sigma_{S^{d+1}}$ \c{Polyakov:2001af}, we will treat \eqr{HHLM} as valid in $\sigma_{\rm AdS}$. Then by marginality, the total OPE coefficient is
\es{}{C_{HHL}& \sim (\D_H)^{\D_L^\M/2}(2-\D_H)^{1-\D_L^\M/2} \\&\sim (-1)^{1-\D_L^\M/2}\,\D_H~.}

To summarize: for charged vertex operators $V_H$ with large worldsheet dimension $\D_H^\M \sim \D_H$, dual to heavy boundary operators $\O_H$, the OPE coefficient $C_{HHL}$ is linear. The result made some assumptions about the structure of the worldsheet theory, and has an undesirable phase factor for generic $\D_L^\M$, so its validity may be limited. It would be nice to find a classical worldsheet proof of this statement using sigma model methods or the techniques of \c{costa, zar, Bajnok:2014sza}. 

\ssec{Chiral algebra corollary}

By way of \c{Beem:2013sza, Beem6d}, \eqr{linear} implies the following property of structure constants of 2d chiral algebras:

\begin{quotation} \noindent {\it {\bf Chiral algebra corollary:} Consider a sequence of vertex operator algebras (VOAs) associated (in the sense of \c{Beem:2013sza, Beem6d}) with a sequence of SCFTs admitting a large $c$, large $\D_\gap$ limit. Suppose the VOA contains strong generators of holomorphic weights $h$ and $H$, where}
\end{quotation}
\e{chiralregime}{h \sim \O(1)~, \quad 1\ll H\ll |c_{2d}|^{\#> 0}~.}
\begin{quotation} \noindent {\it where $c_{2d}$ is the VOA central charge. Then the normalized ``planar'' structure constant $f_{hHH}$ has linear asymptotics,}
\end{quotation}
\e{voalinear}{f_{hHH}\Big|_{H\gg1} \sim {H\o \sqrt{|c_{2d}|}}f(h)~.}
\begin{quotation} \noindent {\it for some function $f(h)$.}
\end{quotation}
\vs
\noindent Following the nomenclature of \c{Beem:2017ooy}, a strong generator cannot be written as a composite of other generators. We have written the conjecture in terms of $|c_{2d}|$ to allow for negative central charges (e.g. in the 4d--2d map, $c_{2d}=-12 c_{4d}$ \c{Beem:2013sza}). 

This statement was motivated by holographic considerations in CFT$_d$, but should ultimately be viewed as a purely 2d property of certain VOAs. We would like to view this as a definite criterion for determining {\it which} VOAs may possibly capture protected operator data of SCFTs with large $c$, large $\D_\gap$ limits, for which such a classification is not yet established even at large $c$ (much less at finite $c$). For example, while VOA data does not depend on marginal couplings, it is not generally clear whether a large $c$ limit of a sequence of SCFTs also has large $\D_\gap$.\foot{An interesting and peculiar class of theories for which this is true is the $(A_1, A_{2n})$ Argyres-Douglas series, whose associated VOA is the Virasoro algebra of the $(2,2n+3)$ minimal model \c{Beem:2017ooy}. At $n\rar\i$, one has $a = c = {n \o 2}$. Since $a=c$ is a necessary but not sufficient condition for large $\D_\gap$, one needs more refined information about the single-trace spectrum, which is lacking at present. The linearity of the central charge in the rank of the gauge group evokes vector model constructions \c{kp, gg} which contain light higher-spin operators. It would be worth identifying the gravity dual of these theories in the $n\rar\i$ limit. We thank Leonardo Rastelli for bringing this example to our attention.} Therefore, checking whether \eqr{voalinear} is satisfied in an infinitely-strongly-generated VOA with $|c_{2d}|\rar\i$ may be a useful practical tool for addressing that question, by suggesting that it {\it should} be associated to an SCFT.

\ssec{On a Sublattice Weak Gravity Conjecture for CFT}\label{secwgc}
Our results give us some perspective on an AdS/CFT version of the Weak Gravity Conjecture (WGC) \c{ArkaniHamed:2006dz}, in particular the Sublattice Weak Gravity Conjecture (sLWGC) \c{Heidenreich:2016aqi}. 

The sLWGC states that in any gauge theory coupled to gravity, a finite index sublattice of the full charge lattice should contain a superextremal particle, where extremality is defined by the charge-to-mass ratio of an extremal black hole carrying the relevant charge. The sLWGC has, to our knowledge, survived all tests in flat space theories of gravity. Following earlier work \c{Nakayama:2015hga}, a translation of the usual WGC bound to AdS$_5$, with mass simply replaced by $\D$, was checked for $\cN=4$ SYM at strong coupling in \c{Heidenreich:2016aqi}, which was shown to satisfy the sLWGC. That inequality is
\e{}{\left({q_\O\o \D_\O}\right)^2 \geq {c_J \o 12 c_T}~,}
where $c_J \sim \la J|J\ra$ is the ``current central charge'', $c_T \sim \la T|T\ra$ is the ordinary central charge, and $q_\O$ is a $\mf{u}(1)$ charge. 

Let us add some comments to this. In order to ponder a CFT version of the WGC at all, one should restrict attention to families of CFTs with limits of large $c$ and large $\D_\gap$, since WGC arguments apply when the gravitational theory contains an Einstein subsector at low-energy. In such CFTs, $c_J \sim c_T$ (up to constants), so the ratio is parametrically bounded simply by a constant. There is no widely accepted generalization of the WGC or sLWGC to AdS quantum gravity which includes $\O(1)$ constants in the charge-to-mass bound and applies to every case.\foot{In \c{Crisford:2017gsb, Horowitz:2019eum}, there is evidence that cosmic censorship suggests the correctness of the literal translation of certain non-lattice forms of WGC to AdS$_4$, though this suggestion is weaker in the presence of dilatonic scalar couplings to Maxwell fields.} To make a connection to extra dimensions, we need only study asymptotic versions of the WGC or sLWGC. Thus, we consider the following parametric formulation: 
\e{caswgc}{ \text{\bf Parametric sLWGC}: \quad {C_2(\R_p) \o \D_p^2}\Big|_{p\gg1} \gtrsim \text{constant}>0}
where $C_2(\R_p)$ is the quadratic Casimir of an irrep $\R_p$ of some global symmetry, furnished by a local operator $\O_p$. $C_2(\R_p)$ becomes large along any direction in the Cartan. This is an sLWGC because the condition is being satisfied asymptotically. It is a coarse condition, not only by way of being parametric, but also because in the presence of multiple $\mf{u}(1)$'s it is weaker than the convex hull condition \c{Cheung:2014vva}. 

What is the connection between  \eqr{caswgc} and extra dimensions? First, in any AdS$\times \M$ compactification where $L_\M\sim L$, \eqr{caswgc} will be satisfied for the symmetries dual to local isometries of $\M$. This is true on account of the asymptotic proportionality $C_2(\R_p) \propto \D_p$ for KK modes $\O_p$, as discussed around \eqr{casimir}. We would like to understand this from the CFT side. But this follows precisely from linearity of $C_{\phi pp}$ at $\D_p\gg1$ in cases where $C_{\phi pp}$ can be identified with charge, e.g. when $\phi$ lives in a supermultiplet of a $\mf{u}(1)$ current. In such cases, $q_p/\D_p$ approaches a constant asymptotically. This proves the parametric sLWGC for all SCFTs with current multiplets containing scalars, and infinite towers of operators of increasing charge.

On the other hand, even this coarse version of the sLWGC is not always satisfied. First, putative CFTs with local AdS$_{d+1}$ duals may not satisfy \eqr{caswgc} because they lack light charge-carrying operators. Another exception is a CFT with baryonic symmetry $\mf{u}(1)^{b_2(\M)}$ arising from topological cycles of manifolds $\M$ with second Betti number $b_2(\M)$: this symmetry also has no light charge carriers, since $\D_{\rm baryon}$ scales as $N$, the rank of the gauge group.\foot{In making this statement, we are understanding the sLWGC as applicable to single-trace towers only, whereas baryonic operators are dual to determinant-type operators (and can undergo multi-trace mixing).} Finally, an exception that {\it does} include light charge-carriers is when the tower $\O_p$ doesn't have charge growing with $p$. This occurs for some flavor symmetries in SCFTs, for which the half-BPS superconformal primaries $\O_p$ are in the adjoint for all $p$, as discussed earlier. The discussion in \c{Palti:2019pca} is consistent with these comments.

\ssec{Linearity of chiral primary three-point functions in D1-D5 CFT}\label{appLM}
We follow \c{Lunin:2001pw}. They compute three-point functions of chiral primary operators in the $\cN=4$ symmetric orbifolds $\text{Sym}_N(M)$ at large $N$, where $M$ is $T^4$ or $K3$. The operators are gauge-invariant chiral twist fields $\O_n^{1_n,\bar 1_n}$, where
\e{}{(h_n,\bar h_n) = \left({n+1_n\o 2},{n+\bar 1_n\o 2}\right)~,~~1_n, \bar 1_n = \pm~.}
For details, see \c{Lunin:2001pw} and \c{Lunin:2000yv}.

At leading order in large $N \sim c$, the general normalized three-point coefficient for these operators is\foot{See eqs. 6.39-6.41 and 6.47 of \c{Lunin:2001pw}. We determined the 3j symbol by charge conservation. Specific examples are given in \c{Lunin:2001pw}.}
\e{LM1}{C_{nmq}^{1_n\bar 1_n;1_m\bar 1_m;1_q\bar 1_q} = {\sqrt{mnq}\o \sqrt{N}}\times |\hat{C}_{nmq}^{1_n;1_m;1_q}|^2\times  \left|\def\arraystretch{1.3}\left(\begin{array}{ccc} {n+1_n\o 2}&{m+1_m\o 2}&{q+1_q\o 2}\\{q-m+1+1_n\o 2}&{m+1_m\o 2}&-\left({q+1+1_n+1_m\o 2}\right)\end{array} \right)\right|^2}
where $\hat{C}_{nmq}^{1_n;1_m;1_q}$ is the ``reduced'' OPE coefficient,
\e{LM2}{\hat{C}_{nmq}^{1_n;1_m;1_q} =  \left({(1_n+n+1_m+m+1_q+q)^2\o 4mnq}{\Sigma!\alpha_n!\alpha_m!\alpha_q!\o (n+1_n)!(m+1_m)!(q+1_q)!}\right)^{1/2}}
with
\es{}{\Sigma &= {1\o 2}(n+m+q+1_n+1_m+1_q)+1\\
\alpha_n &= \Sigma-1_n-1~,}
and the factor in parenthesis in \eqr{LM1} is an $\mf{su}(2)$ 3j symbol. $|\cdot|^2$ denotes multiplication of holomorphic and anti-holomorphic pieces, where, in particular, $1_n$ and $\bar 1_n$ may be chosen independently. We want to take $m=q$ large
for fixed $n$. When $m=q$ the 3j symbol \foot{The 3j symbols are easily computed using the Mathematica command {\tt ThreeJSymbol}.} vanishes unless $1_n=-1$, and takes the form 
\e{}{\left(\begin{array}{ccc} {n-1\o 2}&{q+1_q\o 2}&{q+1_q\o 2}\\0&{q+1_q\o 2}&-\left({q+1_q\o 2}\right)\end{array} \right) = {\Gamma(q+1_q+1)\o \sqrt{\Gamma\left({q+1_q+{3-n\o 2}}\right)\Gamma\left({q+1_q+{3+n\o 2}}\right)}}}
At $q\gg1$, 
\e{}{{\Gamma(q+1_q+1)\o \sqrt{\Gamma\left({q+1_q+{3-n\o 2}}\right)\Gamma\left({q+1_q+{3+n\o 2}}\right)}}\Bigg|_{q\gg1} = {1\o \sqrt{q}}+\O(q^{-3/2})}
The reduced OPE coefficient scales as
\e{}{\hat{C}_{nqq}^{-1;1_q;1_q}\Big|_{q\gg1} = {\Gamma\left({1+n\o 2}\right)\o \sqrt{\Gamma(1+n)}}\sqrt{q}+\O(q^{-1/2})}
Putting it all together, 
\e{}{C_{nqq}^{1_n\bar 1_n;1_m\bar 1_m;1_q\bar 1_q}\Big|_{q\gg1} \sim {q\o \sqrt{N}}{\Gamma\left({1+n\o 2}\right)^2\o \sqrt{n}\,\Gamma(n)}}
This linearity in $q$ is consistent with our claim, because this OPE coefficient is not renormalized along moduli space.  

Compared to the result of \c{Lunin:2000yv} where they perform the computation for $N$ free bosons without SUSY, the OPE data given above is much simpler. However, one can analyze the bosonic result as above, and we find linearity again. The computation is significantly more involved. The bosonic result has no clear relation to our conjecture, which requires large $\D_\gap$ and hence an interpolation in moduli space from an orbifold point, but it would be worthwhile to understand this better.

\bibliographystyle{utphys} 
\bibliography{extradimsbib}

\providecommand{\href}[2]{#2}\begingroup\raggedright\begin{thebibliography}{100}

\bibitem{ps}
J.~Polchinski and E.~Silverstein,
  \href{http://dx.doi.org/10.1142/9789814412551_0018}{``{Dual Purpose
  Landscaping Tools: Small Extra Dimensions in AdS/CFT},''} in {\em Strings,
  gauge fields, and the geometry behind: The legacy of Maximilian Kreuzer},
  A.~Rebhan, L.~Katzarkov, J.~Knapp, R.~Rashkov, and E.~Scheidegger, eds.,
  pp.~365--390.
\newblock 2009.
\newblock \href{http://arxiv.org/abs/0908.0756}{{\ttfamily arXiv:0908.0756
  [hep-th]}}.
\newblock
\url{http://www-public.slac.stanford.edu/sciDoc/docMeta.aspx?slacPubNumber=SLAC-PUB-13748}.
\newblock

\bibitem{hpps}
I.~Heemskerk, J.~Penedones, J.~Polchinski, and J.~Sully, ``{Holography from
  Conformal Field Theory},''
  \href{http://dx.doi.org/10.1088/1126-6708/2009/10/079}{{\em JHEP} {\bfseries
  10} (2009) 079},
\href{http://arxiv.org/abs/0907.0151}{{\ttfamily arXiv:0907.0151 [hep-th]}}.

\bibitem{showkimas}
S.~El-Showk and K.~Papadodimas, ``{Emergent Spacetime and Holographic CFTs},''
  \href{http://dx.doi.org/10.1007/JHEP10(2012)106}{{\em JHEP} {\bfseries 10}
  (2012) 106},
\href{http://arxiv.org/abs/1101.4163}{{\ttfamily arXiv:1101.4163 [hep-th]}}.

\bibitem{banks}
T.~Banks, ``{TASI Lectures on Holographic Space-Time, SUSY and Gravitational
  Effective Field Theory},'' in {\em {Proceedings, Theoretical Advanced Study
  Institute in Elementary Particle Physics (TASI 2010). String Theory and Its
  Applications: From meV to the Planck Scale: Boulder, Colorado, USA, June
  1-25, 2010}}.
\newblock 2010.
\newblock
\href{http://arxiv.org/abs/1007.4001}{{\ttfamily arXiv:1007.4001 [hep-th]}}.
\newblock

\bibitem{fkap}
A.~L. Fitzpatrick and J.~Kaplan, ``{AdS Field Theory from Conformal Field
  Theory},'' \href{http://dx.doi.org/10.1007/JHEP02(2013)054}{{\em JHEP}
  {\bfseries 02} (2013) 054},
\href{http://arxiv.org/abs/1208.0337}{{\ttfamily arXiv:1208.0337 [hep-th]}}.

\bibitem{cemz}
X.~O. Camanho, J.~D. Edelstein, J.~Maldacena, and A.~Zhiboedov, ``{Causality
  Constraints on Corrections to the Graviton Three-Point Coupling},''
  \href{http://dx.doi.org/10.1007/JHEP02(2016)020}{{\em JHEP} {\bfseries 02}
  (2016) 020},
\href{http://arxiv.org/abs/1407.5597}{{\ttfamily arXiv:1407.5597 [hep-th]}}.

\bibitem{Polyakov:2004br}
A.~M. Polyakov, ``{Conformal fixed points of unidentified gauge theories},''
  \href{http://dx.doi.org/10.1142/S0217732304015129}{{\em Mod. Phys. Lett.}
  {\bfseries A19} (2004) 1649--1660},
  \href{http://arxiv.org/abs/hep-th/0405106}{{\ttfamily arXiv:hep-th/0405106
  [hep-th]}}.
[,1159(2004)].

\bibitem{Klebanov:2004ya}
I.~R. Klebanov and J.~M. Maldacena, ``{Superconformal gauge theories and
  non-critical superstrings},''
  \href{http://dx.doi.org/10.1142/S0217751X04020865}{{\em Int. J. Mod. Phys.}
  {\bfseries A19} (2004) 5003--5016},
\href{http://arxiv.org/abs/hep-th/0409133}{{\ttfamily arXiv:hep-th/0409133
  [hep-th]}}.

\bibitem{Kuperstein:2004yk}
S.~Kuperstein and J.~Sonnenschein, ``{Noncritical supergravity (d>1) and
  holography},'' \href{http://dx.doi.org/10.1088/1126-6708/2004/07/049}{{\em
  JHEP} {\bfseries 07} (2004) 049},
\href{http://arxiv.org/abs/hep-th/0403254}{{\ttfamily arXiv:hep-th/0403254
  [hep-th]}}.

\bibitem{Gadde:2009dj}
A.~Gadde, E.~Pomoni, and L.~Rastelli, ``{The Veneziano Limit of N = 2
  Superconformal QCD: Towards the String Dual of N = 2 SU(N(c)) SYM with N(f) =
  2 N(c)},''
\href{http://arxiv.org/abs/0912.4918}{{\ttfamily arXiv:0912.4918 [hep-th]}}.

\bibitem{Douglas:2010ic}
M.~R. Douglas, ``{Spaces of Quantum Field Theories},''
  \href{http://dx.doi.org/10.1088/1742-6596/462/1/012011}{{\em J. Phys. Conf.
  Ser.} {\bfseries 462} no.~1, (2013) 012011},
\href{http://arxiv.org/abs/1005.2779}{{\ttfamily arXiv:1005.2779 [hep-th]}}.

\bibitem{kklt}
S.~Kachru, R.~Kallosh, A.~D. Linde, and S.~P. Trivedi, ``{De Sitter vacua in
  string theory},'' \href{http://dx.doi.org/10.1103/PhysRevD.68.046005}{{\em
  Phys. Rev.} {\bfseries D68} (2003) 046005},
\href{http://arxiv.org/abs/hep-th/0301240}{{\ttfamily arXiv:hep-th/0301240
  [hep-th]}}.

\bibitem{Balasubramanian:2005zx}
V.~Balasubramanian, P.~Berglund, J.~P. Conlon, and F.~Quevedo, ``{Systematics
  of moduli stabilisation in Calabi-Yau flux compactifications},''
  \href{http://dx.doi.org/10.1088/1126-6708/2005/03/007}{{\em JHEP} {\bfseries
  03} (2005) 007},
\href{http://arxiv.org/abs/hep-th/0502058}{{\ttfamily arXiv:hep-th/0502058
  [hep-th]}}.

\bibitem{DeWolfe:2005uu}
O.~DeWolfe, A.~Giryavets, S.~Kachru, and W.~Taylor, ``{Type IIA moduli
  stabilization},'' \href{http://dx.doi.org/10.1088/1126-6708/2005/07/066}{{\em
  JHEP} {\bfseries 07} (2005) 066},
\href{http://arxiv.org/abs/hep-th/0505160}{{\ttfamily arXiv:hep-th/0505160
  [hep-th]}}.

\bibitem{Acharya:2006zw}
B.~S. Acharya and M.~R. Douglas, ``{A Finite landscape?},''
\href{http://arxiv.org/abs/hep-th/0606212}{{\ttfamily arXiv:hep-th/0606212
  [hep-th]}}.

\bibitem{McOrist:2012yc}
J.~McOrist and S.~Sethi, ``{M-theory and Type IIA Flux Compactifications},''
  \href{http://dx.doi.org/10.1007/JHEP12(2012)122}{{\em JHEP} {\bfseries 12}
  (2012) 122},
\href{http://arxiv.org/abs/1208.0261}{{\ttfamily arXiv:1208.0261 [hep-th]}}.

\bibitem{Petrini:2013ika}
M.~Petrini, G.~Solard, and T.~Van~Riet, ``{AdS vacua with scale separation from
  IIB supergravity},'' \href{http://dx.doi.org/10.1007/JHEP11(2013)010}{{\em
  JHEP} {\bfseries 11} (2013) 010},
\href{http://arxiv.org/abs/1308.1265}{{\ttfamily arXiv:1308.1265 [hep-th]}}.

\bibitem{Gautason:2015tig}
F.~F. Gautason, M.~Schillo, T.~Van~Riet, and M.~Williams, ``{Remarks on scale
  separation in flux vacua},''
  \href{http://dx.doi.org/10.1007/JHEP03(2016)061}{{\em JHEP} {\bfseries 03}
  (2016) 061},
\href{http://arxiv.org/abs/1512.00457}{{\ttfamily arXiv:1512.00457 [hep-th]}}.

\bibitem{Sethi:2017phn}
S.~Sethi, ``{Supersymmetry Breaking by Fluxes},''
  \href{http://dx.doi.org/10.1007/JHEP10(2018)022}{{\em JHEP} {\bfseries 10}
  (2018) 022},
\href{http://arxiv.org/abs/1709.03554}{{\ttfamily arXiv:1709.03554 [hep-th]}}.

\bibitem{Alday:2014tsa}
L.~F. Alday, A.~Bissi, and T.~Lukowski, ``{Lessons from crossing symmetry at
  large N},'' \href{http://dx.doi.org/10.1007/JHEP06(2015)074}{{\em JHEP}
  {\bfseries 06} (2015) 074},
\href{http://arxiv.org/abs/1410.4717}{{\ttfamily arXiv:1410.4717 [hep-th]}}.

\bibitem{Alday:2016htq}
L.~F. Alday and A.~Bissi, ``{Unitarity and positivity constraints for CFT at
  large central charge},''
  \href{http://dx.doi.org/10.1007/JHEP07(2017)044}{{\em JHEP} {\bfseries 07}
  (2017) 044},
\href{http://arxiv.org/abs/1606.09593}{{\ttfamily arXiv:1606.09593 [hep-th]}}.

\bibitem{Aharony:2016dwx}
O.~Aharony, L.~F. Alday, A.~Bissi, and E.~Perlmutter, ``{Loops in AdS from
  Conformal Field Theory},''
  \href{http://dx.doi.org/10.1007/JHEP07(2017)036}{{\em JHEP} {\bfseries 07}
  (2017) 036},
\href{http://arxiv.org/abs/1612.03891}{{\ttfamily arXiv:1612.03891 [hep-th]}}.

\bibitem{Li:2017lmh}
D.~Li, D.~Meltzer, and D.~Poland, ``{Conformal Bootstrap in the Regge Limit},''
  \href{http://dx.doi.org/10.1007/JHEP12(2017)013}{{\em JHEP} {\bfseries 12}
  (2017) 013},
\href{http://arxiv.org/abs/1705.03453}{{\ttfamily arXiv:1705.03453 [hep-th]}}.

\bibitem{Alday:2017gde}
L.~F. Alday, A.~Bissi, and E.~Perlmutter, ``{Holographic Reconstruction of AdS
  Exchanges from Crossing Symmetry},''
  \href{http://dx.doi.org/10.1007/JHEP08(2017)147}{{\em JHEP} {\bfseries 08}
  (2017) 147},
\href{http://arxiv.org/abs/1705.02318}{{\ttfamily arXiv:1705.02318 [hep-th]}}.

\bibitem{Giombi:2018vtc}
S.~Giombi, V.~Kirilin, and E.~Perlmutter, ``{Double-Trace Deformations of
  Conformal Correlations},''
  \href{http://dx.doi.org/10.1007/JHEP02(2018)175}{{\em JHEP} {\bfseries 02}
  (2018) 175},
\href{http://arxiv.org/abs/1801.01477}{{\ttfamily arXiv:1801.01477 [hep-th]}}.

\bibitem{Rastelli:2016nze}
L.~Rastelli and X.~Zhou, ``{Mellin amplitudes for $AdS_5\times S^5$},''
  \href{http://dx.doi.org/10.1103/PhysRevLett.118.091602}{{\em Phys. Rev.
  Lett.} {\bfseries 118} no.~9, (2017) 091602},
\href{http://arxiv.org/abs/1608.06624}{{\ttfamily arXiv:1608.06624 [hep-th]}}.

\bibitem{Rastelli:2017udc}
L.~Rastelli and X.~Zhou, ``{How to Succeed at Holographic Correlators Without
  Really Trying},'' \href{http://dx.doi.org/10.1007/JHEP04(2018)014}{{\em JHEP}
  {\bfseries 04} (2018) 014},
\href{http://arxiv.org/abs/1710.05923}{{\ttfamily arXiv:1710.05923 [hep-th]}}.

\bibitem{Rastelli:2017ymc}
L.~Rastelli and X.~Zhou, ``{Holographic Four-Point Functions in the (2, 0)
  Theory},'' \href{http://dx.doi.org/10.1007/JHEP06(2018)087}{{\em JHEP}
  {\bfseries 06} (2018) 087},
\href{http://arxiv.org/abs/1712.02788}{{\ttfamily arXiv:1712.02788 [hep-th]}}.

\bibitem{Chester:2018aca}
S.~M. Chester, S.~S. Pufu, and X.~Yin, ``{The M-Theory S-Matrix From ABJM:
  Beyond 11D Supergravity},''
  \href{http://dx.doi.org/10.1007/JHEP08(2018)115}{{\em JHEP} {\bfseries 08}
  (2018) 115},
\href{http://arxiv.org/abs/1804.00949}{{\ttfamily arXiv:1804.00949 [hep-th]}}.

\bibitem{Chester:2018dga}
S.~M. Chester and E.~Perlmutter, ``{M-Theory Reconstruction from (2,0) CFT and
  the Chiral Algebra Conjecture},''
  \href{http://dx.doi.org/10.1007/JHEP08(2018)116}{{\em JHEP} {\bfseries 08}
  (2018) 116},
\href{http://arxiv.org/abs/1805.00892}{{\ttfamily arXiv:1805.00892 [hep-th]}}.

\bibitem{Binder:2018yvd}
D.~J. Binder, S.~M. Chester, and S.~S. Pufu, ``{Absence of $D^4 R^4$ in
  M-Theory From ABJM},''
\href{http://arxiv.org/abs/1808.10554}{{\ttfamily arXiv:1808.10554 [hep-th]}}.

\bibitem{Caron-Huot:2018kta}
S.~Caron-Huot and A.-K. Trinh, ``{All tree-level correlators in AdS$_{5}\times
  S_{5}$ supergravity: hidden ten-dimensional conformal symmetry},''
  \href{http://dx.doi.org/10.1007/JHEP01(2019)196}{{\em JHEP} {\bfseries 01}
  (2019) 196},
\href{http://arxiv.org/abs/1809.09173}{{\ttfamily arXiv:1809.09173 [hep-th]}}.

\bibitem{Binder:2019jwn}
D.~J. Binder, S.~M. Chester, S.~S. Pufu, and Y.~Wang, ``{$\mathcal{N}=4$
  Super-Yang-Mills Correlators at Strong Coupling from String Theory and
  Localization},''
\href{http://arxiv.org/abs/1902.06263}{{\ttfamily arXiv:1902.06263 [hep-th]}}.

\bibitem{Rastelli:2019gtj}
L.~Rastelli, K.~Roumpedakis, and X.~Zhou, ``{$\mathbf{AdS_3\times S^3}$
  Tree-Level Correlators: Hidden Six-Dimensional Conformal Symmetry},''
\href{http://arxiv.org/abs/1905.11983}{{\ttfamily arXiv:1905.11983 [hep-th]}}.

\bibitem{Giusto}
S.~Giusto, R.~Russo, A.~Tyukov, and C.~Wen, ``{Holographic correlators in
  AdS$_3$ without Witten diagrams},''
  \href{http://dx.doi.org/10.1007/JHEP09(2019)030}{{\em JHEP} {\bfseries 09}
  (2019) 030},
\href{http://arxiv.org/abs/1905.12314}{{\ttfamily arXiv:1905.12314 [hep-th]}}.

\bibitem{Alday:2017vkk}
L.~F. Alday and S.~Caron-Huot, ``{Gravitational S-matrix from CFT dispersion
  relations},'' \href{http://dx.doi.org/10.1007/JHEP12(2018)017}{{\em JHEP}
  {\bfseries 12} (2018) 017},
\href{http://arxiv.org/abs/1711.02031}{{\ttfamily arXiv:1711.02031 [hep-th]}}.

\bibitem{Aprile:2017bgs}
F.~Aprile, J.~M. Drummond, P.~Heslop, and H.~Paul, ``{Quantum Gravity from
  Conformal Field Theory},''
  \href{http://dx.doi.org/10.1007/JHEP01(2018)035}{{\em JHEP} {\bfseries 01}
  (2018) 035},
\href{http://arxiv.org/abs/1706.02822}{{\ttfamily arXiv:1706.02822 [hep-th]}}.

\bibitem{Alday:2017xua}
L.~F. Alday and A.~Bissi, ``{Loop Corrections to Supergravity on $AdS_5 \times
  S^5$},'' \href{http://dx.doi.org/10.1103/PhysRevLett.119.171601}{{\em Phys.
  Rev. Lett.} {\bfseries 119} no.~17, (2017) 171601},
\href{http://arxiv.org/abs/1706.02388}{{\ttfamily arXiv:1706.02388 [hep-th]}}.

\bibitem{Aprile:2017xsp}
F.~Aprile, J.~M. Drummond, P.~Heslop, and H.~Paul, ``{Unmixing Supergravity},''
  \href{http://dx.doi.org/10.1007/JHEP02(2018)133}{{\em JHEP} {\bfseries 02}
  (2018) 133},
\href{http://arxiv.org/abs/1706.08456}{{\ttfamily arXiv:1706.08456 [hep-th]}}.

\bibitem{Aprile:2017qoy}
F.~Aprile, J.~M. Drummond, P.~Heslop, and H.~Paul, ``{Loop corrections for
  Kaluza-Klein AdS amplitudes},''
  \href{http://dx.doi.org/10.1007/JHEP05(2018)056}{{\em JHEP} {\bfseries 05}
  (2018) 056},
\href{http://arxiv.org/abs/1711.03903}{{\ttfamily arXiv:1711.03903 [hep-th]}}.

\bibitem{Aprile:2018efk}
F.~Aprile, J.~Drummond, P.~Heslop, and H.~Paul, ``{Double-trace spectrum of
  $N=4$ supersymmetric Yang-Mills theory at strong coupling},''
  \href{http://dx.doi.org/10.1103/PhysRevD.98.126008}{{\em Phys. Rev.}
  {\bfseries D98} no.~12, (2018) 126008},
\href{http://arxiv.org/abs/1802.06889}{{\ttfamily arXiv:1802.06889 [hep-th]}}.

\bibitem{Alday:2018pdi}
L.~F. Alday, A.~Bissi, and E.~Perlmutter, ``{Genus-One String Amplitudes from
  Conformal Field Theory},''
\href{http://arxiv.org/abs/1809.10670}{{\ttfamily arXiv:1809.10670 [hep-th]}}.

\bibitem{Alday:2018kkw}
L.~F. Alday, ``{On Genus-one String Amplitudes on $AdS_5 \times S^5$},''
\href{http://arxiv.org/abs/1812.11783}{{\ttfamily arXiv:1812.11783 [hep-th]}}.

\bibitem{Ponomarev:2019ltz}
D.~Ponomarev, E.~Sezgin, and E.~Skvortsov, ``{On one loop corrections in higher
  spin gravity},''
\href{http://arxiv.org/abs/1904.01042}{{\ttfamily arXiv:1904.01042 [hep-th]}}.

\bibitem{LSPT}
L.~F. Alday, ``{Large Spin Perturbation Theory for Conformal Field Theories},''
  \href{http://dx.doi.org/10.1103/PhysRevLett.119.111601}{{\em Phys. Rev.
  Lett.} {\bfseries 119} no.~11, (2017) 111601},
\href{http://arxiv.org/abs/1611.01500}{{\ttfamily arXiv:1611.01500 [hep-th]}}.

\bibitem{simon}
S.~Caron-Huot, ``{Analyticity in Spin in Conformal Theories},''
  \href{http://dx.doi.org/10.1007/JHEP09(2017)078}{{\em JHEP} {\bfseries 09}
  (2017) 078},
\href{http://arxiv.org/abs/1703.00278}{{\ttfamily arXiv:1703.00278 [hep-th]}}.

\bibitem{Gary:2009ae}
M.~Gary, S.~B. Giddings, and J.~Penedones, ``{Local bulk S-matrix elements and
  CFT singularities},''
  \href{http://dx.doi.org/10.1103/PhysRevD.80.085005}{{\em Phys. Rev.}
  {\bfseries D80} (2009) 085005},
\href{http://arxiv.org/abs/0903.4437}{{\ttfamily arXiv:0903.4437 [hep-th]}}.

\bibitem{pen}
J.~Penedones, ``{Writing CFT correlation functions as AdS scattering
  amplitudes},'' \href{http://dx.doi.org/10.1007/JHEP03(2011)025}{{\em JHEP}
  {\bfseries 03} (2011) 025},
\href{http://arxiv.org/abs/1011.1485}{{\ttfamily arXiv:1011.1485 [hep-th]}}.

\bibitem{Maldacena:2015iua}
J.~Maldacena, D.~Simmons-Duffin, and A.~Zhiboedov, ``{Looking for a bulk
  point},'' \href{http://dx.doi.org/10.1007/JHEP01(2017)013}{{\em JHEP}
  {\bfseries 01} (2017) 013},
\href{http://arxiv.org/abs/1509.03612}{{\ttfamily arXiv:1509.03612 [hep-th]}}.

\bibitem{Hijano:2019qmi}
E.~Hijano, ``{Flat space physics from AdS/CFT},'' {\em Submitted to: J. High
  Energy Phys.} (2019) ,
\href{http://arxiv.org/abs/1905.02729}{{\ttfamily arXiv:1905.02729 [hep-th]}}.

\bibitem{ov}
H.~Ooguri and C.~Vafa, ``{Non-supersymmetric AdS and the Swampland},''
  \href{http://dx.doi.org/10.4310/ATMP.2017.v21.n7.a8}{{\em Adv. Theor. Math.
  Phys.} {\bfseries 21} (2017) 1787--1801},
\href{http://arxiv.org/abs/1610.01533}{{\ttfamily arXiv:1610.01533 [hep-th]}}.

\bibitem{fk}
B.~Freivogel and M.~Kleban, ``{Vacua Morghulis},''
\href{http://arxiv.org/abs/1610.04564}{{\ttfamily arXiv:1610.04564 [hep-th]}}.

\bibitem{ho1}
D.~Harlow and H.~Ooguri, ``{Constraints on symmetry from holography},''
  \href{http://dx.doi.org/10.1103/PhysRevLett.122.191601}{{\em Phys. Rev.
  Lett.} {\bfseries 122} no.~19, (2019) 191601},
\href{http://arxiv.org/abs/1810.05337}{{\ttfamily arXiv:1810.05337 [hep-th]}}.

\bibitem{ho2}
D.~Harlow and H.~Ooguri, ``{Symmetries in quantum field theory and quantum
  gravity},''
\href{http://arxiv.org/abs/1810.05338}{{\ttfamily arXiv:1810.05338 [hep-th]}}.

\bibitem{Cornalba:2007zb}
L.~Cornalba, M.~S. Costa, and J.~Penedones, ``{Eikonal approximation in
  AdS/CFT: Resumming the gravitational loop expansion},''
  \href{http://dx.doi.org/10.1088/1126-6708/2007/09/037}{{\em JHEP} {\bfseries
  09} (2007) 037},
\href{http://arxiv.org/abs/0707.0120}{{\ttfamily arXiv:0707.0120 [hep-th]}}.

\bibitem{star1}
C.~Sleight and M.~Taronna, ``{Spinning Mellin Bootstrap: Conformal Partial
  Waves, Crossing Kernels and Applications},''
  \href{http://dx.doi.org/10.1002/prop.201800038}{{\em Fortsch. Phys.}
  {\bfseries 66} no.~8-9, (2018) 1800038},
\href{http://arxiv.org/abs/1804.09334}{{\ttfamily arXiv:1804.09334 [hep-th]}}.

\bibitem{star2}
C.~Sleight and M.~Taronna, ``{Anomalous Dimensions from Crossing Kernels},''
  \href{http://dx.doi.org/10.1007/JHEP11(2018)089}{{\em JHEP} {\bfseries 11}
  (2018) 089},
\href{http://arxiv.org/abs/1807.05941}{{\ttfamily arXiv:1807.05941 [hep-th]}}.

\bibitem{cppr}
M.~S. Costa, J.~Penedones, D.~Poland, and S.~Rychkov, ``{Spinning Conformal
  Correlators},'' \href{http://dx.doi.org/10.1007/JHEP11(2011)071}{{\em JHEP}
  {\bfseries 11} (2011) 071},
\href{http://arxiv.org/abs/1107.3554}{{\ttfamily arXiv:1107.3554 [hep-th]}}.

\bibitem{costa}
M.~S. Costa, R.~Monteiro, J.~E. Santos, and D.~Zoakos, ``{On three-point
  correlation functions in the gauge/gravity duality},''
  \href{http://dx.doi.org/10.1007/JHEP11(2010)141}{{\em JHEP} {\bfseries 11}
  (2010) 141},
\href{http://arxiv.org/abs/1008.1070}{{\ttfamily arXiv:1008.1070 [hep-th]}}.

\bibitem{abjm}
O.~Aharony, O.~Bergman, D.~L. Jafferis, and J.~Maldacena, ``{N=6 superconformal
  Chern-Simons-matter theories, M2-branes and their gravity duals},''
  \href{http://dx.doi.org/10.1088/1126-6708/2008/10/091}{{\em JHEP} {\bfseries
  10} (2008) 091},
\href{http://arxiv.org/abs/0806.1218}{{\ttfamily arXiv:0806.1218 [hep-th]}}.

\bibitem{Shimizu:2017kzs}
H.~Shimizu, Y.~Tachikawa, and G.~Zafrir, ``{Anomaly matching on the Higgs
  branch},'' \href{http://dx.doi.org/10.1007/JHEP12(2017)127}{{\em JHEP}
  {\bfseries 12} (2017) 127},
\href{http://arxiv.org/abs/1703.01013}{{\ttfamily arXiv:1703.01013 [hep-th]}}.

\bibitem{cdi}
C.~Cordova, T.~T. Dumitrescu, and K.~Intriligator, ``{Multiplets of
  Superconformal Symmetry in Diverse Dimensions},''
  \href{http://dx.doi.org/10.1007/JHEP03(2019)163}{{\em JHEP} {\bfseries 03}
  (2019) 163},
\href{http://arxiv.org/abs/1612.00809}{{\ttfamily arXiv:1612.00809 [hep-th]}}.

\bibitem{Baggio:2012rr}
M.~Baggio, J.~de~Boer, and K.~Papadodimas, ``{A non-renormalization theorem for
  chiral primary 3-point functions},''
  \href{http://dx.doi.org/10.1007/JHEP07(2012)137}{{\em JHEP} {\bfseries 07}
  (2012) 137},
\href{http://arxiv.org/abs/1203.1036}{{\ttfamily arXiv:1203.1036 [hep-th]}}.

\bibitem{Beem:2013sza}
C.~Beem, M.~Lemos, P.~Liendo, W.~Peelaers, L.~Rastelli, and B.~C. van Rees,
  ``{Infinite Chiral Symmetry in Four Dimensions},''
  \href{http://dx.doi.org/10.1007/s00220-014-2272-x}{{\em Commun. Math. Phys.}
  {\bfseries 336} no.~3, (2015) 1359--1433},
\href{http://arxiv.org/abs/1312.5344}{{\ttfamily arXiv:1312.5344 [hep-th]}}.

\bibitem{katz}
A.~L. Fitzpatrick, E.~Katz, D.~Poland, and D.~Simmons-Duffin, ``{Effective
  Conformal Theory and the Flat-Space Limit of AdS},''
  \href{http://dx.doi.org/10.1007/JHEP07(2011)023}{{\em JHEP} {\bfseries 07}
  (2011) 023},
\href{http://arxiv.org/abs/1007.2412}{{\ttfamily arXiv:1007.2412 [hep-th]}}.

\bibitem{DHoker:1999kzh}
E.~D'Hoker, D.~Z. Freedman, S.~D. Mathur, A.~Matusis, and L.~Rastelli,
  ``{Graviton exchange and complete four point functions in the AdS / CFT
  correspondence},''
  \href{http://dx.doi.org/10.1016/S0550-3213(99)00525-8}{{\em Nucl. Phys.}
  {\bfseries B562} (1999) 353--394},
\href{http://arxiv.org/abs/hep-th/9903196}{{\ttfamily arXiv:hep-th/9903196
  [hep-th]}}.

\bibitem{Silverstein:2001xn}
E.~Silverstein, ``{(A)dS backgrounds from asymmetric orientifolds},'' {\em Clay
  Mat. Proc.} {\bfseries 1} (2002) 179,
\href{http://arxiv.org/abs/hep-th/0106209}{{\ttfamily arXiv:hep-th/0106209
  [hep-th]}}.

\bibitem{Aharony:2006ra}
O.~Aharony and E.~Silverstein, ``{Supercritical stability, transitions and
  (pseudo)tachyons},'' \href{http://dx.doi.org/10.1103/PhysRevD.75.046003}{{\em
  Phys. Rev.} {\bfseries D75} (2007) 046003},
\href{http://arxiv.org/abs/hep-th/0612031}{{\ttfamily arXiv:hep-th/0612031
  [hep-th]}}.

\bibitem{Hellerman:2006nx}
S.~Hellerman and I.~Swanson, ``{Cosmological solutions of supercritical string
  theory},'' \href{http://dx.doi.org/10.1103/PhysRevD.77.126011}{{\em Phys.
  Rev.} {\bfseries D77} (2008) 126011},
\href{http://arxiv.org/abs/hep-th/0611317}{{\ttfamily arXiv:hep-th/0611317
  [hep-th]}}.

\bibitem{Green:2007tr}
D.~R. Green, A.~Lawrence, J.~McGreevy, D.~R. Morrison, and E.~Silverstein,
  ``{Dimensional duality},''
  \href{http://dx.doi.org/10.1103/PhysRevD.76.066004}{{\em Phys. Rev.}
  {\bfseries D76} (2007) 066004},
\href{http://arxiv.org/abs/0705.0550}{{\ttfamily arXiv:0705.0550 [hep-th]}}.

\bibitem{hks}
T.~Hartman, C.~A. Keller, and B.~Stoica, ``{Universal Spectrum of 2d Conformal
  Field Theory in the Large c Limit},''
  \href{http://dx.doi.org/10.1007/JHEP09(2014)118}{{\em JHEP} {\bfseries 09}
  (2014) 118},
\href{http://arxiv.org/abs/1405.5137}{{\ttfamily arXiv:1405.5137 [hep-th]}}.

\bibitem{Hellerman:2009bu}
S.~Hellerman, ``{A Universal Inequality for CFT and Quantum Gravity},''
  \href{http://dx.doi.org/10.1007/JHEP08(2011)130}{{\em JHEP} {\bfseries 08}
  (2011) 130},
\href{http://arxiv.org/abs/0902.2790}{{\ttfamily arXiv:0902.2790 [hep-th]}}.

\bibitem{Friedan:2013cba}
D.~Friedan and C.~A. Keller, ``{Constraints on 2d CFT partition functions},''
  \href{http://dx.doi.org/10.1007/JHEP10(2013)180}{{\em JHEP} {\bfseries 10}
  (2013) 180},
\href{http://arxiv.org/abs/1307.6562}{{\ttfamily arXiv:1307.6562 [hep-th]}}.

\bibitem{Collier:2016cls}
S.~Collier, Y.-H. Lin, and X.~Yin, ``{Modular Bootstrap Revisited},''
  \href{http://dx.doi.org/10.1007/JHEP09(2018)061}{{\em JHEP} {\bfseries 09}
  (2018) 061},
\href{http://arxiv.org/abs/1608.06241}{{\ttfamily arXiv:1608.06241 [hep-th]}}.

\bibitem{Afkhami-Jeddi:2019zci}
N.~Afkhami-Jeddi, T.~Hartman, and A.~Tajdini, ``{Fast Conformal Bootstrap and
  Constraints on 3d Gravity},''
  \href{http://dx.doi.org/10.1007/JHEP05(2019)087}{{\em JHEP} {\bfseries 05}
  (2019) 087},
\href{http://arxiv.org/abs/1903.06272}{{\ttfamily arXiv:1903.06272 [hep-th]}}.

\bibitem{Hartman:2019pcd}
T.~Hartman, D.~Maz\'a\v{c}, and L.~Rastelli, ``{Sphere Packing and Quantum
  Gravity},''
\href{http://arxiv.org/abs/1905.01319}{{\ttfamily arXiv:1905.01319 [hep-th]}}.

\bibitem{Haehl:2014yla}
F.~M. Haehl and M.~Rangamani, ``{Permutation orbifolds and holography},''
  \href{http://dx.doi.org/10.1007/JHEP03(2015)163}{{\em JHEP} {\bfseries 03}
  (2015) 163},
\href{http://arxiv.org/abs/1412.2759}{{\ttfamily arXiv:1412.2759 [hep-th]}}.

\bibitem{Belin:2014fna}
A.~Belin, C.~A. Keller, and A.~Maloney, ``{String Universality for Permutation
  Orbifolds},'' \href{http://dx.doi.org/10.1103/PhysRevD.91.106005}{{\em Phys.
  Rev.} {\bfseries D91} no.~10, (2015) 106005},
\href{http://arxiv.org/abs/1412.7159}{{\ttfamily arXiv:1412.7159 [hep-th]}}.

\bibitem{Henningson:1998gx}
M.~Henningson and K.~Skenderis, ``{The Holographic Weyl anomaly},''
  \href{http://dx.doi.org/10.1088/1126-6708/1998/07/023}{{\em JHEP} {\bfseries
  07} (1998) 023},
\href{http://arxiv.org/abs/hep-th/9806087}{{\ttfamily arXiv:hep-th/9806087
  [hep-th]}}.

\bibitem{dyer}
E.~Dyer~(unpublished).

\bibitem{Lashkari:2016vgj}
N.~Lashkari, A.~Dymarsky, and H.~Liu, ``{Eigenstate Thermalization Hypothesis
  in Conformal Field Theory},''
  \href{http://dx.doi.org/10.1088/1742-5468/aab020}{{\em J. Stat. Mech.}
  {\bfseries 1803} no.~3, (2018) 033101},
\href{http://arxiv.org/abs/1610.00302}{{\ttfamily arXiv:1610.00302 [hep-th]}}.

\bibitem{km}
P.~Kraus and A.~Maloney, ``{A cardy formula for three-point coefficients or how
  the black hole got its spots},''
  \href{http://dx.doi.org/10.1007/JHEP05(2017)160}{{\em JHEP} {\bfseries 05}
  (2017) 160},
\href{http://arxiv.org/abs/1608.03284}{{\ttfamily arXiv:1608.03284 [hep-th]}}.

\bibitem{Hellerman:2015nra}
S.~Hellerman, D.~Orlando, S.~Reffert, and M.~Watanabe, ``{On the CFT Operator
  Spectrum at Large Global Charge},''
  \href{http://dx.doi.org/10.1007/JHEP12(2015)071}{{\em JHEP} {\bfseries 12}
  (2015) 071},
\href{http://arxiv.org/abs/1505.01537}{{\ttfamily arXiv:1505.01537 [hep-th]}}.

\bibitem{datta}
D.~Das, S.~Datta, and S.~Pal, ``{Charged structure constants from
  modularity},'' \href{http://dx.doi.org/10.1007/JHEP11(2017)183}{{\em JHEP}
  {\bfseries 11} (2017) 183},
\href{http://arxiv.org/abs/1706.04612}{{\ttfamily arXiv:1706.04612 [hep-th]}}.

\bibitem{Lee:1998bxa}
S.~Lee, S.~Minwalla, M.~Rangamani, and N.~Seiberg, ``{Three point functions of
  chiral operators in D = 4, N=4 SYM at large N},''
  \href{http://dx.doi.org/10.4310/ATMP.1998.v2.n4.a1}{{\em Adv. Theor. Math.
  Phys.} {\bfseries 2} (1998) 697--718},
\href{http://arxiv.org/abs/hep-th/9806074}{{\ttfamily arXiv:hep-th/9806074
  [hep-th]}}.

\bibitem{Bastianelli:1999en}
F.~Bastianelli and R.~Zucchini, ``{Three point functions of chiral primary
  operators in d = 3, N=8 and d = 6, N=(2,0) SCFT at large N},''
  \href{http://dx.doi.org/10.1016/S0370-2693(99)01179-X}{{\em Phys. Lett.}
  {\bfseries B467} (1999) 61--66},
\href{http://arxiv.org/abs/hep-th/9907047}{{\ttfamily arXiv:hep-th/9907047
  [hep-th]}}.

\bibitem{Lunin:2001pw}
O.~Lunin and S.~D. Mathur, ``{Three point functions for M(N) / S(N) orbifolds
  with N=4 supersymmetry},''
  \href{http://dx.doi.org/10.1007/s002200200638}{{\em Commun. Math. Phys.}
  {\bfseries 227} (2002) 385--419},
\href{http://arxiv.org/abs/hep-th/0103169}{{\ttfamily arXiv:hep-th/0103169
  [hep-th]}}.

\bibitem{Hirano:2012vz}
S.~Hirano, C.~Kristjansen, and D.~Young, ``{Giant Gravitons on AdS$_4 x CP^3$
  and their Holographic Three-point Functions},''
  \href{http://dx.doi.org/10.1007/JHEP07(2012)006}{{\em JHEP} {\bfseries 07}
  (2012) 006},
\href{http://arxiv.org/abs/1205.1959}{{\ttfamily arXiv:1205.1959 [hep-th]}}.

\bibitem{bk}
A.~Bissi, C.~Kristjansen, A.~Martirosyan, and M.~Orselli, ``{On Three-point
  Functions in the AdS$_4$/CFT$_3$ Correspondence},''
  \href{http://dx.doi.org/10.1007/JHEP01(2013)137}{{\em JHEP} {\bfseries 01}
  (2013) 137},
\href{http://arxiv.org/abs/1211.1359}{{\ttfamily arXiv:1211.1359 [hep-th]}}.

\bibitem{zar}
K.~Zarembo, ``{Holographic three-point functions of semiclassical states},''
  \href{http://dx.doi.org/10.1007/JHEP09(2010)030}{{\em JHEP} {\bfseries 09}
  (2010) 030},
\href{http://arxiv.org/abs/1008.1059}{{\ttfamily arXiv:1008.1059 [hep-th]}}.

\bibitem{mp}
J.~A. Minahan and R.~Pereira, ``{Three-point correlators from string
  amplitudes: Mixing and Regge spins},''
  \href{http://dx.doi.org/10.1007/JHEP04(2015)134}{{\em JHEP} {\bfseries 04}
  (2015) 134},
\href{http://arxiv.org/abs/1410.4746}{{\ttfamily arXiv:1410.4746 [hep-th]}}.

\bibitem{roiban}
R.~Roiban and A.~A. Tseytlin, ``{On semiclassical computation of 3-point
  functions of closed string vertex operators in $AdS_5 x S^5$},''
  \href{http://dx.doi.org/10.1103/PhysRevD.82.106011}{{\em Phys. Rev.}
  {\bfseries D82} (2010) 106011},
\href{http://arxiv.org/abs/1008.4921}{{\ttfamily arXiv:1008.4921 [hep-th]}}.

\bibitem{bmp}
T.~Bargheer, J.~A. Minahan, and R.~Pereira, ``{Computing Three-Point Functions
  for Short Operators},'' \href{http://dx.doi.org/10.1007/JHEP03(2014)096}{{\em
  JHEP} {\bfseries 03} (2014) 096},
\href{http://arxiv.org/abs/1311.7461}{{\ttfamily arXiv:1311.7461 [hep-th]}}.

\bibitem{Baggio:2016skg}
M.~Baggio, V.~Niarchos, K.~Papadodimas, and G.~Vos, ``{Large-N correlation
  functions in $ \mathcal{N} $ = 2 superconformal QCD},''
  \href{http://dx.doi.org/10.1007/JHEP01(2017)101}{{\em JHEP} {\bfseries 01}
  (2017) 101},
\href{http://arxiv.org/abs/1610.07612}{{\ttfamily arXiv:1610.07612 [hep-th]}}.

\bibitem{bmn}
D.~E. Berenstein, J.~M. Maldacena, and H.~S. Nastase, ``{Strings in flat space
  and pp waves from N=4 superYang-Mills},''
  \href{http://dx.doi.org/10.1088/1126-6708/2002/04/013}{{\em JHEP} {\bfseries
  04} (2002) 013},
\href{http://arxiv.org/abs/hep-th/0202021}{{\ttfamily arXiv:hep-th/0202021
  [hep-th]}}.

\bibitem{Hellerman:2017veg}
S.~Hellerman, S.~Maeda, and M.~Watanabe, ``{Operator Dimensions from Moduli},''
  \href{http://dx.doi.org/10.1007/JHEP10(2017)089}{{\em JHEP} {\bfseries 10}
  (2017) 089},
\href{http://arxiv.org/abs/1706.05743}{{\ttfamily arXiv:1706.05743 [hep-th]}}.

\bibitem{joetalk}
J.~Polchinski, ``{Landscape/CFT Duality?},'' {\em KITP seminar} (December 11,
  2008) .
  \url{http://online.kitp.ucsb.edu/online/joint98/polchinski4/rm/jwvideo.html}.

\bibitem{Klebanov:1998hh}
I.~R. Klebanov and E.~Witten, ``{Superconformal field theory on three-branes at
  a Calabi-Yau singularity},''
  \href{http://dx.doi.org/10.1016/S0550-3213(98)00654-3}{{\em Nucl. Phys.}
  {\bfseries B536} (1998) 199--218},
\href{http://arxiv.org/abs/hep-th/9807080}{{\ttfamily arXiv:hep-th/9807080
  [hep-th]}}.

\bibitem{gubser}
S.~S. Gubser, ``{Einstein manifolds and conformal field theories},''
  \href{http://dx.doi.org/10.1103/PhysRevD.59.025006}{{\em Phys. Rev.}
  {\bfseries D59} (1999) 025006},
\href{http://arxiv.org/abs/hep-th/9807164}{{\ttfamily arXiv:hep-th/9807164
  [hep-th]}}.

\bibitem{ferrara1}
A.~Ceresole, G.~Dall'Agata, R.~D'Auria, and S.~Ferrara, ``{Superconformal field
  theories from IIB spectroscopy on AdS(5) x T-11},''
  \href{http://dx.doi.org/10.1088/0264-9381/17/5/311}{{\em Class. Quant. Grav.}
  {\bfseries 17} (2000) 1017--1025},
\href{http://arxiv.org/abs/hep-th/9910066}{{\ttfamily arXiv:hep-th/9910066
  [hep-th]}}.

\bibitem{ferrara2}
A.~Ceresole, G.~Dall'Agata, R.~D'Auria, and S.~Ferrara, ``{Spectrum of type IIB
  supergravity on AdS(5) x T**11: Predictions on N=1 SCFT's},''
  \href{http://dx.doi.org/10.1103/PhysRevD.61.066001}{{\em Phys. Rev.}
  {\bfseries D61} (2000) 066001},
\href{http://arxiv.org/abs/hep-th/9905226}{{\ttfamily arXiv:hep-th/9905226
  [hep-th]}}.

\bibitem{Cvetic:2005ft}
M.~Cvetic, H.~Lu, D.~N. Page, and C.~N. Pope, ``{New Einstein-Sasaki spaces in
  five and higher dimensions},''
  \href{http://dx.doi.org/10.1103/PhysRevLett.95.071101}{{\em Phys. Rev. Lett.}
  {\bfseries 95} (2005) 071101},
\href{http://arxiv.org/abs/hep-th/0504225}{{\ttfamily arXiv:hep-th/0504225
  [hep-th]}}.

\bibitem{Sparks:2010sn}
J.~Sparks, ``{Sasaki-Einstein Manifolds},''
  \href{http://dx.doi.org/10.4310/SDG.2011.v16.n1.a6}{{\em Surveys Diff. Geom.}
  {\bfseries 16} (2011) 265--324},
\href{http://arxiv.org/abs/1004.2461}{{\ttfamily arXiv:1004.2461 [math.DG]}}.

\bibitem{fabbri}
D.~Fabbri, P.~Fre, L.~Gualtieri, and P.~Termonia, ``{M theory on AdS(4) x
  M**111: The Complete Osp(2|4) x SU(3) x SU(2) spectrum from harmonic
  analysis},'' \href{http://dx.doi.org/10.1016/S0550-3213(99)00363-6}{{\em
  Nucl. Phys.} {\bfseries B560} (1999) 617--682},
\href{http://arxiv.org/abs/hep-th/9903036}{{\ttfamily arXiv:hep-th/9903036
  [hep-th]}}.

\bibitem{page1}
D.~N. Page and C.~N. Pope, ``{Which Compactifications of $D=11$ Supergravity
  Are Stable?},''
\href{http://dx.doi.org/10.1016/0370-2693(84)91275-9}{{\em Phys. Lett.}
  {\bfseries 144B} (1984) 346--350}.

\bibitem{page2}
D.~N. Page and C.~N. Pope, ``{Stability Analysis of Compactifications of $D=11$
  Supergravity With SU(3) X SU(2) X U(1) Symmetry},''
\href{http://dx.doi.org/10.1016/0370-2693(84)90056-X}{{\em Phys. Lett.}
  {\bfseries 145B} (1984) 337--341}.

\bibitem{yasuda}
O.~Yasuda, ``Classical stability of ${M}^{\mathrm{pqr}}$, ${Q}^{\mathrm{pqr}}$,
  and ${N}^{\mathrm{pqr}}$ in $d=11$ supergravity,''
  \href{http://dx.doi.org/10.1103/PhysRevLett.53.1207}{{\em Phys. Rev. Lett.}
  {\bfseries 53} (Sep, 1984) 1207--1211}.
  \url{https://link.aps.org/doi/10.1103/PhysRevLett.53.1207}.

\bibitem{Dong:2015gya}
X.~Dong, D.~Z. Freedman, and Y.~Zhao, ``{AdS/CFT and the Little Hierarchy
  Problem},''
\href{http://arxiv.org/abs/1510.01741}{{\ttfamily arXiv:1510.01741 [hep-th]}}.

\bibitem{banksov}
T.~Banks, ``{Note on a Paper by Ooguri and Vafa},''
\href{http://arxiv.org/abs/1611.08953}{{\ttfamily arXiv:1611.08953 [hep-th]}}.

\bibitem{Giombi:2017mxl}
S.~Giombi and E.~Perlmutter, ``{Double-Trace Flows and the Swampland},''
  \href{http://dx.doi.org/10.1007/JHEP03(2018)026}{{\em JHEP} {\bfseries 03}
  (2018) 026},
\href{http://arxiv.org/abs/1709.09159}{{\ttfamily arXiv:1709.09159 [hep-th]}}.

\bibitem{besse}
A.~Besse, {\em Einstein Manifolds}.
\newblock Classics in mathematics. Springer, 1987.
\newblock \url{https://books.google.com/books?id=6I\_XgRJaBL0C}.

\bibitem{anderson}
M.~T. {Anderson}, ``{A survey of Einstein metrics on 4-manifolds},'' {\em arXiv
  e-prints} (Oct, 2008) arXiv:0810.4830,
  \href{http://arxiv.org/abs/0810.4830}{{\ttfamily arXiv:0810.4830 [math.DG]}}.

\bibitem{yang}
D.~Yang, ``{Rigidity of Einstein 4-manifolds with positive curvature},''
  \href{http://dx.doi.org/10.1007/PL00005792}{{\em Inventiones mathematicae}
  {\bfseries 142} no.~2, (Nov, 2000) 435--450}.
  \url{https://doi.org/10.1007/PL00005792}.

\bibitem{Buchel:2008vz}
A.~Buchel, R.~C. Myers, and A.~Sinha, ``{Beyond eta/s = 1/4 pi},''
  \href{http://dx.doi.org/10.1088/1126-6708/2009/03/084}{{\em JHEP} {\bfseries
  03} (2009) 084},
\href{http://arxiv.org/abs/0812.2521}{{\ttfamily arXiv:0812.2521 [hep-th]}}.

\bibitem{Bhardwaj:2013qia}
L.~Bhardwaj and Y.~Tachikawa, ``{Classification of 4d N=2 gauge theories},''
  \href{http://dx.doi.org/10.1007/JHEP12(2013)100}{{\em JHEP} {\bfseries 12}
  (2013) 100},
\href{http://arxiv.org/abs/1309.5160}{{\ttfamily arXiv:1309.5160 [hep-th]}}.

\bibitem{Polyakov:2001af}
A.~M. Polyakov, ``{Gauge fields and space-time},''
  \href{http://dx.doi.org/10.1142/S0217751X02013071}{{\em Int. J. Mod. Phys.}
  {\bfseries A17S1} (2002) 119--136},
\href{http://arxiv.org/abs/hep-th/0110196}{{\ttfamily arXiv:hep-th/0110196
  [hep-th]}}.

\bibitem{Tseytlin:2003ac}
A.~A. Tseytlin, ``{On semiclassical approximation and spinning string vertex
  operators in AdS(5) x S**5},''
  \href{http://dx.doi.org/10.1016/S0550-3213(03)00456-5}{{\em Nucl. Phys.}
  {\bfseries B664} (2003) 247--275},
\href{http://arxiv.org/abs/hep-th/0304139}{{\ttfamily arXiv:hep-th/0304139
  [hep-th]}}.

\bibitem{Bajnok:2014sza}
Z.~Bajnok, R.~A. Janik, and A.~Wereszczynski, ``{HHL correlators, orbit
  averaging and form factors},''
  \href{http://dx.doi.org/10.1007/JHEP09(2014)050}{{\em JHEP} {\bfseries 09}
  (2014) 050},
\href{http://arxiv.org/abs/1404.4556}{{\ttfamily arXiv:1404.4556 [hep-th]}}.

\bibitem{Beem6d}
C.~Beem, L.~Rastelli, and B.~C. van Rees, ``{$ \mathcal{W} $ symmetry in six
  dimensions},'' \href{http://dx.doi.org/10.1007/JHEP05(2015)017}{{\em JHEP}
  {\bfseries 05} (2015) 017},
\href{http://arxiv.org/abs/1404.1079}{{\ttfamily arXiv:1404.1079 [hep-th]}}.

\bibitem{Beem:2017ooy}
C.~Beem and L.~Rastelli, ``{Vertex operator algebras, Higgs branches, and
  modular differential equations},''
  \href{http://dx.doi.org/10.1007/JHEP08(2018)114}{{\em JHEP} {\bfseries 08}
  (2018) 114},
\href{http://arxiv.org/abs/1707.07679}{{\ttfamily arXiv:1707.07679 [hep-th]}}.

\bibitem{kp}
I.~R. Klebanov and A.~M. Polyakov, ``{AdS dual of the critical O(N) vector
  model},'' \href{http://dx.doi.org/10.1016/S0370-2693(02)02980-5}{{\em Phys.
  Lett.} {\bfseries B550} (2002) 213--219},
\href{http://arxiv.org/abs/hep-th/0210114}{{\ttfamily arXiv:hep-th/0210114
  [hep-th]}}.

\bibitem{gg}
M.~R. Gaberdiel and R.~Gopakumar, ``{An AdS$_3$ Dual for Minimal Model CFTs},''
  \href{http://dx.doi.org/10.1103/PhysRevD.83.066007}{{\em Phys. Rev.}
  {\bfseries D83} (2011) 066007},
\href{http://arxiv.org/abs/1011.2986}{{\ttfamily arXiv:1011.2986 [hep-th]}}.

\bibitem{ArkaniHamed:2006dz}
N.~Arkani-Hamed, L.~Motl, A.~Nicolis, and C.~Vafa, ``{The String landscape,
  black holes and gravity as the weakest force},''
  \href{http://dx.doi.org/10.1088/1126-6708/2007/06/060}{{\em JHEP} {\bfseries
  06} (2007) 060},
\href{http://arxiv.org/abs/hep-th/0601001}{{\ttfamily arXiv:hep-th/0601001
  [hep-th]}}.

\bibitem{Heidenreich:2016aqi}
B.~Heidenreich, M.~Reece, and T.~Rudelius, ``{Evidence for a sublattice weak
  gravity conjecture},'' \href{http://dx.doi.org/10.1007/JHEP08(2017)025}{{\em
  JHEP} {\bfseries 08} (2017) 025},
\href{http://arxiv.org/abs/1606.08437}{{\ttfamily arXiv:1606.08437 [hep-th]}}.

\bibitem{Nakayama:2015hga}
Y.~Nakayama and Y.~Nomura, ``{Weak gravity conjecture in the AdS/CFT
  correspondence},'' \href{http://dx.doi.org/10.1103/PhysRevD.92.126006}{{\em
  Phys. Rev.} {\bfseries D92} no.~12, (2015) 126006},
\href{http://arxiv.org/abs/1509.01647}{{\ttfamily arXiv:1509.01647 [hep-th]}}.

\bibitem{Crisford:2017gsb}
T.~Crisford, G.~T. Horowitz, and J.~E. Santos, ``{Testing the Weak Gravity -
  Cosmic Censorship Connection},''
  \href{http://dx.doi.org/10.1103/PhysRevD.97.066005}{{\em Phys. Rev.}
  {\bfseries D97} no.~6, (2018) 066005},
\href{http://arxiv.org/abs/1709.07880}{{\ttfamily arXiv:1709.07880 [hep-th]}}.

\bibitem{Horowitz:2019eum}
G.~T. Horowitz and J.~E. Santos, ``{Further evidence for the weak gravity -
  cosmic censorship connection},''
\href{http://arxiv.org/abs/1901.11096}{{\ttfamily arXiv:1901.11096 [hep-th]}}.

\bibitem{Cheung:2014vva}
C.~Cheung and G.~N. Remmen, ``{Naturalness and the Weak Gravity Conjecture},''
  \href{http://dx.doi.org/10.1103/PhysRevLett.113.051601}{{\em Phys. Rev.
  Lett.} {\bfseries 113} (2014) 051601},
\href{http://arxiv.org/abs/1402.2287}{{\ttfamily arXiv:1402.2287 [hep-ph]}}.

\bibitem{Palti:2019pca}
E.~Palti, ``{The Swampland: Introduction and Review},''
\newblock 2019.
\newblock
\href{http://arxiv.org/abs/1903.06239}{{\ttfamily arXiv:1903.06239 [hep-th]}}.
\newblock

\bibitem{Lunin:2000yv}
O.~Lunin and S.~D. Mathur, ``{Correlation functions for M**N / S(N)
  orbifolds},'' \href{http://dx.doi.org/10.1007/s002200100431}{{\em Commun.
  Math. Phys.} {\bfseries 219} (2001) 399--442},
\href{http://arxiv.org/abs/hep-th/0006196}{{\ttfamily arXiv:hep-th/0006196
  [hep-th]}}.

\end{thebibliography}\endgroup

\end{document}